\documentclass[aps,showpacs,amsmath,twocolumn,showkeys,pra,superscriptaddress]{revtex4-1}
\usepackage{dcolumn}    
\usepackage{ifpdf}
\usepackage{amssymb,lineno,amsfonts}
\usepackage{graphicx}   
\usepackage{dcolumn}    
\usepackage{bm}         
\usepackage{bbm}
\usepackage{mathrsfs}
\usepackage{upgreek}
\usepackage{setspace}
\usepackage{hyperref}
\usepackage{float}
\usepackage[usenames,dvipsnames]{xcolor}
\definecolor{med-blue}{RGB}{25,25,112}
\hypersetup{colorlinks, linkcolor={red},citecolor={blue}, urlcolor={MidnightBlue}}

\newcommand{\ket}[1]{\vert{#1}\rangle}
\newcommand{\bra}[1]{\langle{#1}\vert}

\newcommand{\expv}[1]{\langle{#1}\rangle}

\newcommand{\expec}[1]{\langle{#1}\rangle}

\begin{document}
\title{Ancilla assisted measurements on quantum ensembles: \\
General protocols and applications in NMR quantum information processing}
\author{T. S. Mahesh}
\def\andname{,}
\email{mahesh.ts@iiserpune.ac.in}
\author{Abhishek Shukla$ ^1 $, Swathi S. Hegde$ ^1 $, \\
C. S. Sudheer Kumar$ ^1 $, Hemant Katiyar$ ^1 $, Sharad Joshi}
\address{Department of Physics and NMR Research Center,\\
	Indian Institute of Science Education and Research, Pune 411008, India}
\author{K. R. Koteswara Rao}
\address{Department of Physics and CQIQC,
Indian Institute of Science, Bangalore 560012, India}

\begin{abstract}
{
Quantum ensembles form easily accessible architectures for studying
various phenomena in quantum physics, quantum information science, and spectroscopy.
Here we review some recent protocols for measurements in quantum ensembles
by utilizing ancillary systems.  We also illustrate these protocols experimentally via nuclear magnetic resonance techniques.  In particular, we shall review 
noninvasive measurements, extracting expectation values of various operators, characterizations of quantum states, and quantum processes, and finally quantum noise engineering.
}
\end{abstract}

\maketitle

\section{Introduction}
Unlike the classical measurements, measurements in quantum physics
affect the dynamics of the system.  Moreover, often a particular experimental
technique may offer only a limited set of observables which can be directly
measured.  The complete characterization of a quantum state or a quantum process
requires, in general, a series of measurements of noncommuting observables - requiring repeated state preparation, and a large number independent measurements.
In this article, we review recent progresses in the measurement of
quantum ensembles and explain how to overcome the above challenges.  
Here we exploit the presence of an \textit{ancillary} register interacting
with the \textit{system} that is to be measured.  In the following section we show  how to realize noninvasive measurements using ancillary qubits.  Extracting
expectation values of various types of operators and related applications are 
described in section III.  We then describe an efficient protocols for 
complete quantum state characterization (section IV) and quantum process tomography
(section V).  We also narrate our experiments on noise engineering using ancillary qubits in section VI, and finally we summarize all the topics in section VII.
In all the sections, we illustrate the protocols experimentally using nuclear magnetic resonance (NMR) techniques.

\section{Noninvasive measurements}
A classical measurement can in principle be \textit{noninvasive} 
in the sense it has no effect on the dynamics of the system.
The same is not true in general for a quantum system, wherein 
the process of measurement itself may affect the dynamics of the system.
However, as explained below, ancillary qubits can be utilized to realize
certain quantum measurements without much disturbance, and hence extract  probabilities or expectation values noninvasively to a great extent.  Such
measurements are often termed as \textit{noninvasive quantum measurements}.

\begin{figure}
\begin{center}
\includegraphics[trim= 5cm 8cm 7cm 5cm,clip=true,width=8cm]{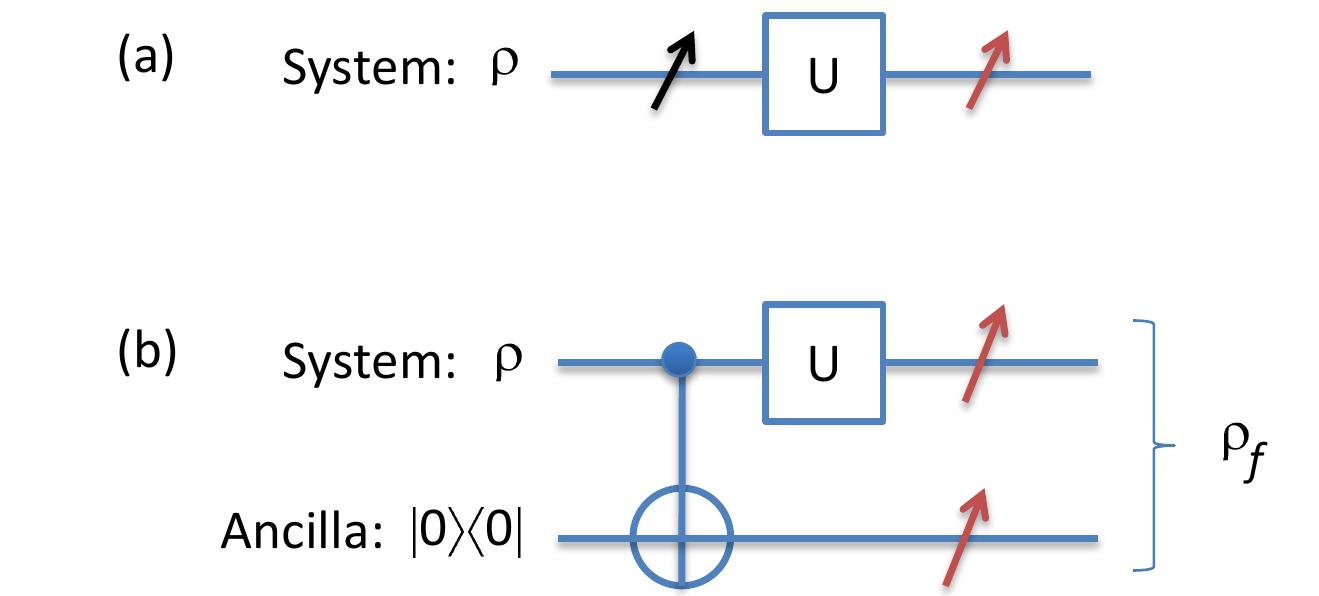}
\caption{Circuits describing invasive and non-invasive measurements.
The first measurement is invasive in (a).  The same is replaced by 
a noninvasive measurement in (b).
The final measurements are invasive in both.}
\label{nim}
\end{center}
\end{figure}

Consider the example of a single qubit initialized in state $\rho$ (Fig. \ref{nim} (a)) and a dichotomic observable $Q$ with eigenvalues $q \in \{0,1\}$. 
Suppose we need to extract the joint probabilities $P(q_1,q_2)$ at two time-instances, one before applying an unitary $U$ and the other after applying $U$ (Fig. \ref{nim} (a)).
To realize the first measurement noninvasively, we utilize an \textit{ancilla} qubit
(Fig. \ref{nim} (b)), a {\tt CNOT} gate, and a final 2-qubit projective measurement in $Q\otimes Q$ basis.  The {\tt CNOT} operation copies the probabilities of $P(q_1)$ onto the ancilla qubit without projecting the system state.  Denoting the first qubit as the ancilla and the second qubit as the system, the diagonal elements of the joint density operator $\rho_f$ store the
joint probabilities, i.e.,
$P(q_1,q_2) = \bra{q_1q_2}\rho_f \ket{q_1q_2}$.
The joint probabilities can thus be extracted by a final strong measurement or
by a diagonal density matrix tomography \cite{elgimahesh}. 
Knee et al \cite{knee} argued that since {\tt CNOT} operator flips the ancilla qubit if the system qubit is in state $\ket{1}$, the circuit is not quite noninvasive.  They also proposed a simple variation, in which $P(0,0)$ and $P(0,1)$ are measured using {\tt CNOT} operator, while $P(1,0)$ and $P(1,1)$ are measured using an
{\tt Anti-CNOT} operator (which flips the ancilla only if the system qubit is in state $\ket{1}$).  In this procedure, called \textit{ideal negative-result measurement} (INRM), all the joint probabilities are measured without flipping the ancilla qubit, and therefore is considered \textit{more noninvasive}.  In the following we describe the application of INRM in studying Leggett-Garg inequalities.

Leggett-Garg inequality (LGI) provides one way of distinguishing quantum behaviour from \textit{macrorealism}  
Macrorealism is based on the following assumptions:
(i) the object remains in one or the other of many possible states at all times, and
(ii) the measurements are noninvasive, i.e., they reveal the state of the object without disturbing the object or its future dynamics.
Leggett-Garg inequality (LGI) sets up macrorealistic bounds 
on linear combinations of two-time correlations of a dichotomic observable
belonging to a single dynamical system \cite{lgi1985}.
Quantum systems are incompatible with these criteria and often violate
bounds on correlations derived from them, thereby allowing us to
distinguish the quantum behavior from macrorealism.
Violations of LGI by quantum systems have been investigated and 
demonstrated experimentally in various systems
\cite{palacios2010experimental,athalye2011investigation,Lambert2011,goggin2011violation,jordanPRL2011,elgimahesh,Oliveira2011timeinequality,etrans,suzuki2012violation,violationQuantumdots,zhou2012experimental,knee2012violation}.
An entropic formulation of LGI has also been introduced  
by Usha Devi \textit{et al.} \cite{elgiUshadevi} 
in terms of classical Shannon entropies associated 
with classical correlations.  
We had reported an experimental demonstration of violation of entropic LGI (ELGI) 
in an ensemble of spin $1/2$ nuclei using NMR techniques \cite{elgimahesh}.
The simplest ELGI study involves three sets of two-time joint measurements of 
a dynamic observable belonging to a `system' qubit at time instants
$(t_1,t_2)$, $(t_2,t_3)$, and $(t_1,t_3)$.  The first measurement in each
case must be noninvasive, and can be performed with the help of an ancilla qubit.

Usha Devi \textit{et al.} \cite{elgiUshadevi} have shown theoretically that for $n$-equidistant measurements on a spin-$s$ system, the information deficit,
\cite{elgiUshadevi}
\begin{equation}
D_n(\theta)=\frac{(n-1)H[\theta / (n-1)] - H[ \theta ]}{\log_2(2s+1)} \geq 0.
\end{equation}
Here the conditional entropies $H(\theta)$ are obtained by the conditional probabilities
\begin{eqnarray}
H\left(
\frac{m\theta}{n-1}
\right) && = H(Q_n \vert Q_{n-m}) \nonumber \\
= && -\sum_{q_{n-m} , q_n}P(q_n \vert q_{n-m}) \log_2P(q_n \vert q_{n-m}), \nonumber
\end{eqnarray}
where $m \in \{1,\cdots, n\}$.
The conditional probabilities in turn are calculated from the joint probabilities using Bayes theorem,
\begin{eqnarray}
P(q_j\vert q_k)P(q_k)=P(q_j,q_k).
\end{eqnarray}

\begin{figure}
\begin{center}
\includegraphics[trim= 6cm 4cm 5cm 4cm,clip=true,width=8cm]{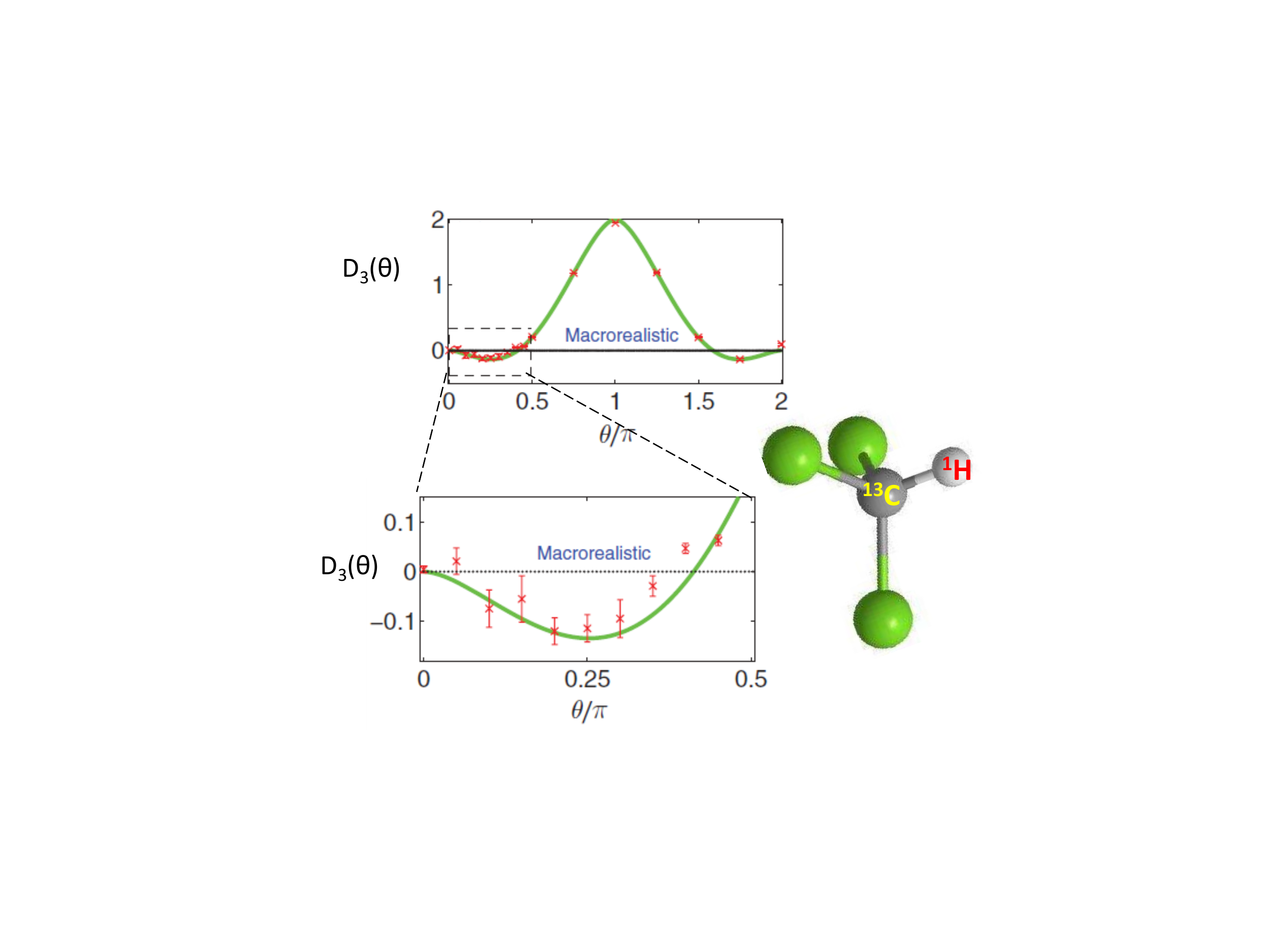}
\caption{Information deficit $D_3$ versus $\theta$ obtained
using INRM procedure.
The mean experimental D3 values are shown as symbols. The curves indicate theoretical D3. The horizontal lines at D3 = 0 indicate the lower bounds of the macrorealism territories
(Figure reproduced from reference \cite{elgimahesh}).}
\label{elgi}
\end{center}
\end{figure}

We studied ELGI experimentally by
treating the $^{13}$C and $^1$H nuclear spins of
$^{13}$CHCl$_3$ (dissolved in CDCl$_3$)
as the system and the ancilla qubits respectively 
(Fig. \ref{elgi}).
The resonance offset of $^{13}$C was set to 100 Hz and
that of $^1$H to 0 Hz (on resonant).  The two spins
have an indirect spin-spin coupling constant $J=209.2$ Hz.
The NMR experiments were carried out at an ambient temperature
of 300 K on a 500 MHz Bruker NMR spectrometer.

As described in earlier,
two sets of experiments were performed, one with {\tt CNOT} and 
the other with {\tt anti-CNOT} \cite{elgimahesh}.
The joint entropies were calculated using the
experimental probabilities and 
the information deficit (in bits) was calculated using the expression
$D_3 = 2H(Q_2 \vert Q_1) - H(Q_3 \vert Q_1)$.
The theoretical and experimental values of $D_3$ for various rotation
angles $\theta$ are shown in Fig. \ref{elgi}.  
According to quantum theory, a maximum violation of $D_3 = -0.134$ should occur 
at $\theta = \pi/4$.  The corresponding experimental value, $D_3(\pi/4) = -0.114 \pm 0.027$, indicates a clear violation of ELGI.

Our other experiments involving noninvasive measurements include
(i) illustrating the inconsistency of quantum marginal probabilities 
with classical probability theory \cite{elgimahesh} and 
(ii) demonstrating that quantum joint probabilities can not be obtained from 
moment distribution \cite{MI}.

\section{Extracting expectation values}
\begin{figure}
\begin{center}
\includegraphics[trim= 0cm 0cm 5cm 0cm,clip=true,width=9.4cm]{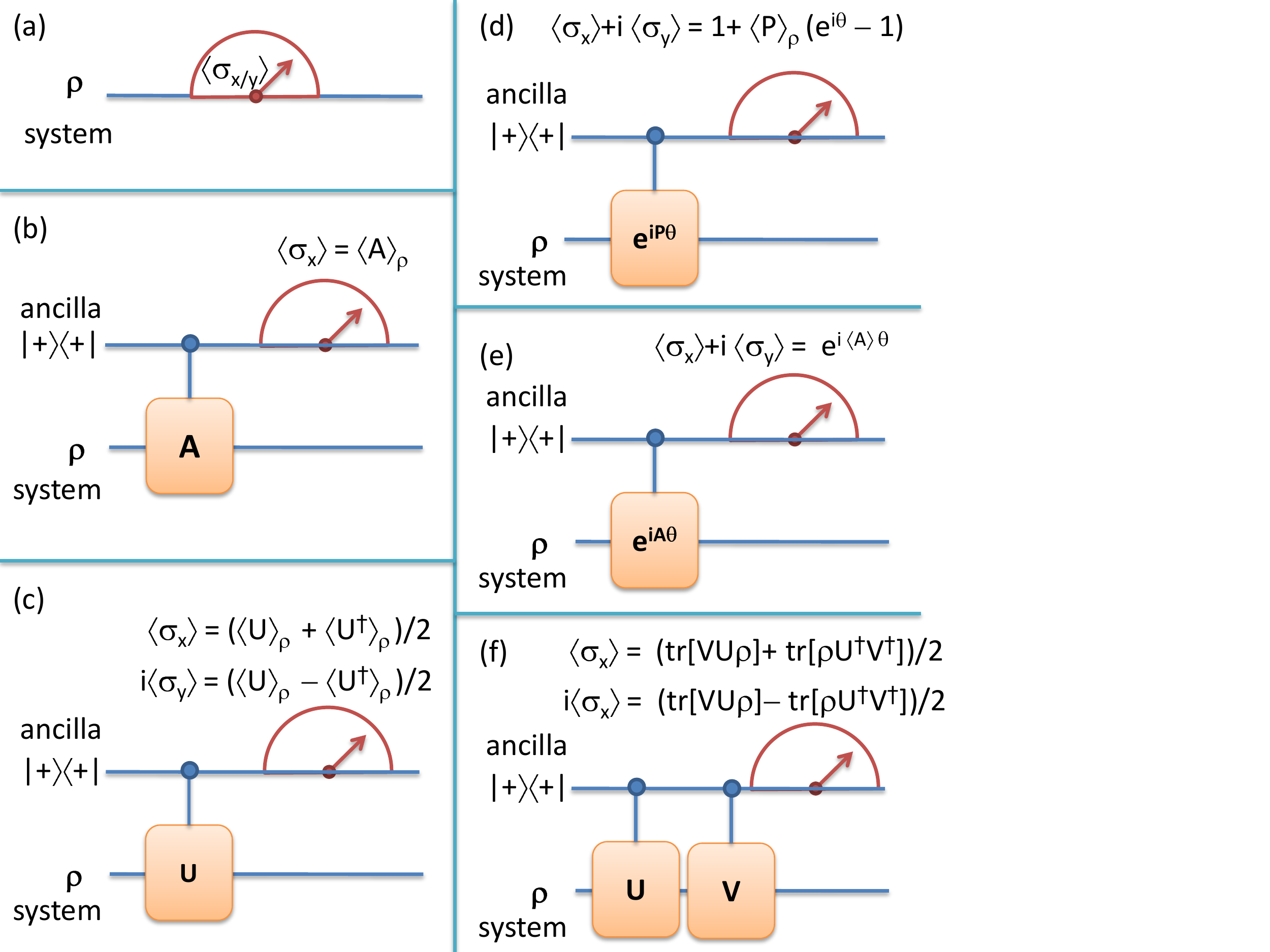}
\caption{Circuits for (a) NMR measurements of $\expv{\sigma_x}$ and $\expv{\sigma_y}$, (b) standard Moussa protocol for expectation values of Hermitian-unitary operator $A$ \cite{moussa1qbitcontextuality2010}, and circuits for measuring expectation values of
(c) a unitary operator $U$ \cite{joshi2014estimating},
(d) a projector $P$ \cite{joshi2014estimating}
(e) a diagonal Hermitian operator $A$, and
(f) for measuring joint expectation values of noncommuting unitaries 
$U$ and $V$\cite{MI}.
}
\label{moussa}
\end{center}
\end{figure}

Often experimental setups allow direct detection of only a limited
set of observables and to extract their expectation values.
For example, in NMR only transverse magnetization operators ($\expv{\sigma_x}$ and $\expv{\sigma_y})$ are directly observable via real and imaginary components of
induced emf.
Fig. \ref{moussa} describes the circuits for measuring expectation
values of different types of operators using an ancilla qubit.
Here the expectation values $\expv{\sigma_x}$ and $\expv{\sigma_y})$
of ancilla qubit reveal the expectation values of different types of
operators acting on system qubit.
Applications of such circuits are illustrated with the help of following experiments:
(A) estimation of Franck-Condon factors, and
(B) investigation of quantum contextuality.  

\subsection{Estimation of Franck-Condon factors}
Franck-Condon principle states that
the transition probability between two vibronic levels depends on the 
overlap between the respective vibrational wavefunctions \cite{demtroder2006atoms}.  
Franck-Condon factors (FCFs) dictate the intensities of vibronic transitions, and therefore their estimation is an important task in understanding 
absorption and fluorescence spectra and related phenomena such as photo-induced dissociations \cite{bowers1984fragmentation}.  

We modelled the electronic ground and excited vibrational levels as eigenstates of two harmonic potentials,
$V_1$ and $V_2$ respectively.  To simulate the one-dimensional case, we choose the potentials $V_1 = x^2/2$ and $V_2 = (x-b)^2/2 + \Delta E$, 
which are identical up to an overall displacement $b$ in position and/or in energy $\Delta E$. 

Thus the vibrational Hamiltonians for the two electronic states are
\begin{eqnarray}
{\cal H}_1 &=& \textbf{p}^2/2 + \textbf{x}^2/2 \nonumber \\
{\cal H}_2 &=& \textbf{p}^2/2 + (\textbf{x}-b)^2/2 +  \Delta E.
\end{eqnarray}
The FCF between $\ket{m}$ of the electronic ground state and $\ket{n}$ 
of the electronic excited state is given by \cite{demtroder2006atoms},
\begin{eqnarray}
f_{m,n'}(b) &=& \vert \langle m \vert n' \rangle \vert^2 \nonumber \\
&=&
\begin{array}{|c|}
\int\limits_{-\infty}^{\infty} \psi_m^*(x) \psi_{n'}(x,b) dx  
\end{array}^2,
\end{eqnarray}
where $\psi_m(x)$, $\psi_{n'}(x,b)$ are the corresponding position wave-functions.

Estimation of FCF, $f_{m,n'}$, is equivalent to measurement of expectation
value of the projection $P_m = \vert m \rangle \langle m \vert$
after preparing the system in excited state $\vert n' \rangle$ since
\cite{joshi2014estimating},
\begin{eqnarray}
f_{m,n'} &=& \langle n' \vert m \rangle \langle m \vert n' \rangle \nonumber \\
&=& \langle P_m \rangle_{n'}.
\label{fcexp}
\end{eqnarray}

\begin{figure}
\centering
\includegraphics[trim=0cm 0cm 0cm 0cm, clip=true,width=8.5cm]{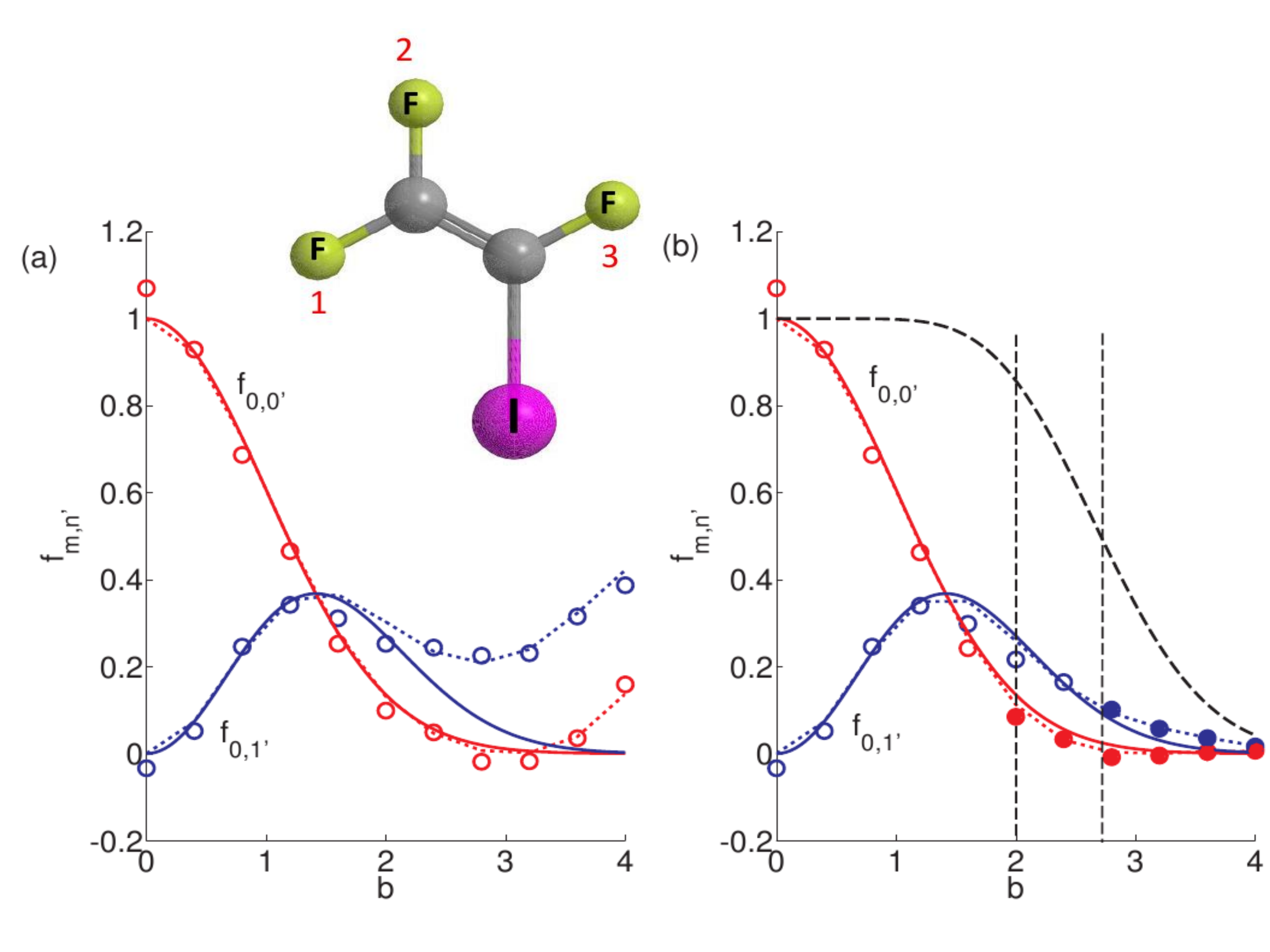}
\caption{Experimental FCFs (circles) corresponding to 4-level harmonic oscillators 
versus the displacement $b$ (in atomic units).
The simulated FCFs (dotted lines) for the 4-level system and
analytical FCFs (smooth lines) for infinite-level system are also shown 
for comparison.
The dashed curve at the top of (b) corresponds to the
normalization used.  The thin vertical dashed lines at $b=2,~1+\sqrt{3}$
mark the beginning of classically forbidden regions for $f_{0,0'},
f_{0,1'}$ respectively.  Molecular structure and qubit labelling are shown in 
the inset (Figure reproduced from reference \cite{joshi2014estimating}).
} 
\label{fcresults}
\end{figure}

Three spin-1/2 $^{19}$F nuclei of iodotrifluoroethylene
(C$_2$F$_3$I) dissolved in acetone-D$_6$ form a three-qubit NMR quantum simulator
(see Fig. \ref{fcresults}).
$F_1$ qubit is chosen to be the
ancilla and the other two qubits chosen representing the lowest four levels
of the Harmonic oscillator \cite{joshi2014estimating}.  
The vibrational levels of the electronic ground
state are encoded onto the spin states such that
$\ket{0} = \ket{\uparrow \uparrow}$, $\ket{1} = \ket{\uparrow \downarrow}$, 
$\ket{2} = \ket{\downarrow \uparrow}$, and
$\ket{3} = \ket{\downarrow \downarrow}$.
The preparation of excited state $\vert n' \rangle$ can be achieved by 
first initializing the system
in the corresponding  state $\vert n \rangle$ of the electronic
ground state and translating it in position from origin ($x=0$)
to the point $x=b$.  
This translation was be achieved by the unitary operator
\begin{eqnarray}
U_T(b) = e^{-i \textbf{p} b}.
\end{eqnarray}
Finally the expectation values $\expv{P_m}$ were measured experimentally
using the circuit shown in Fig. \ref{moussa}d, and then 
the FCFs $f_{m,n}$ were obtained using eqn. \ref{fcexp} \cite{joshi2014estimating}.
The results described in Fig. \ref{fcresults} display a good 
correspondence with the theoretically expected values indicating
the success of the experimental protocols.

\subsection{Investigation of quantum contextuality in a harmonic oscillator} 
Quantum contextuality (QC) states that the outcome of the measurement depends
not only on the system and the observable but also on the 
context of the measurement, i.e., on other compatible observables which are measured along with \cite{peres_context_1pg}.
Consider the following NCHV inequality \cite{quant_info_neilson_chuang}:
\begin{eqnarray}
\mathrm{\textbf{I}}~ && =\expec{AB+BC+CD-AD} \nonumber \\
&& = \expec{AB} + \expec{BC} + \expec{CD} - \expec{AD} \leq 2
\label{I_main}
\end{eqnarray}

Hong-Yi Su \textit{et.al.} \cite{Cont_theory} theoretically studied QC of eigenstates of $1$D-QHO.
They introduced two sets of pseudo-spin operators,
\begin{eqnarray}
\Gamma_x = \sigma_x \otimes \mathbbm{1},~
\Gamma_y = \sigma_z \otimes \sigma_y,~
\Gamma_z = -\sigma_y \otimes \sigma_y,  \nonumber \\
\Gamma_x' = \sigma_x \otimes \sigma_z,~
\Gamma_y' = \mathbbm{1} \otimes \sigma_y,~
\Gamma_z' = -\sigma_x \otimes \sigma_x.  
\label{Gammas defined}
\end{eqnarray}
where, $ \mathbbm{1} $ is $2\times 2$ Identity matrix. Using these operators they defined the observables,
\begin{eqnarray}
A=\Gamma_x,B=c_\beta \Gamma_x' + s_\beta \Gamma_z'=\sigma_x\otimes(c_\beta \sigma_z - s_\beta \sigma_x)  \nonumber \\
C=\Gamma_z, ~~D=c_\eta \Gamma_x' + s_\eta \Gamma_z'=\sigma_x\otimes(c_\eta \sigma_z - s_\eta \sigma_x) \label{A,B,C,D defined},
\label{AB,BC,..evalted}
\end{eqnarray}
where 
$c_\beta = \cos \beta,~s_\beta = \sin \beta, ~ 
c_\eta = \cos \eta$, and $s_\eta = \sin \eta$.
Here operators $A,~B,~C,~D$ are unitary \& Hermitian and accordingly have eigenvalues $ \pm 1 $, with $ (A,B)$, $(B,C)$, $(C,D)$ and $(D,A) $ forming compatible pairs. Hong-Yi Su \textit{et.al.} have shown that,
\begin{eqnarray}
\mathrm{\textbf{I}}_{\ket{l}_{QHO}}^{QM} = 2\sqrt{2} > 2 ,~\mathrm{when},~(\beta,\eta)_l = 
\begin{cases}
(-\pi/4 , -3\pi/4)_0 \\
(3\pi/4, \pi/4)_1 \\
(\pi/4, 3\pi/4)_2 \\
(-3\pi/4, -\pi/4)_3 
\end{cases} 
\label{max vio angles}
\end{eqnarray}
where, $\mathrm{\textbf{I}}_{\ket{l}_{QHO}}^{QM}$ is the expression on LHS of inequality \ref{I_main}, $ l=0,1,2\textrm{ and } 3 $, and, $ \ket{0}_{QHO}, \ket{1}_{QHO}, \ket{2}_{QHO}$ and $ \ket{3}_{QHO}$ are first four energy eigenstates of 1D-QHO. 

\begin{figure}
\begin{center}
\includegraphics[trim= 0cm 4.5cm 0cm 2.2cm, clip=true,width=8.8 cm]{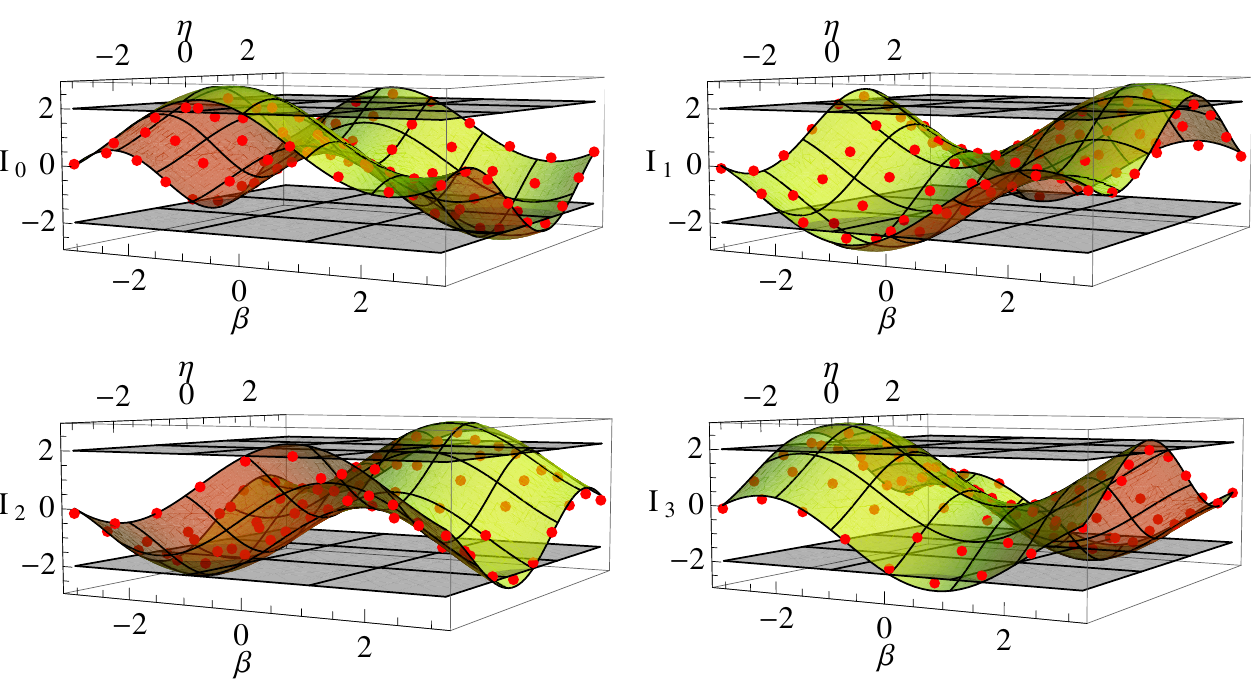}
\caption{$ \mathrm{\textbf{I}}_0 $, $ \mathrm{\textbf{I}}_1 $, $\mathrm{\textbf{I}}_2 $, and $ \mathrm{\textbf{I}}_3 $ represent evaluation of inequality \ref{I_main} for eigenstates $ \ket{0}_{QHO},\ket{1}_{QHO},\ket{2}_{QHO}$, and $\ket{3}_{QHO} $
respectively \cite{katiyar2015investigation}. The curved surface represents theoretical values, and the points are experimental values. Flat planes at $I = 2 $ and $I= -2 $ represent classical bounds.}
\label{context}
\end{center}
\end{figure}

We encoded the first four energy eigenstates of 1D-QHO onto the four Zeeman eigenstates of a pair of spin-1/2 nuclei. The circuit shown in Fig. \ref{moussa}f was used to extract the expectation value of observables ($AB, BC, CD,$ and $DA$) in a joint measurement.  We used three $^{19}$F nuclear spins of trifluoroiodoethylene dissolved in acetone-D6 (see inset of Fig. \ref{fcresults}) as the 3-qubit register. 
The first spin, F$_1$, was used as an ancilla qubit, and other spins, F$_2$ and F$_3$, as the system qubits. The results are shown in Fig. \ref{context}.
The maximum theoretical violation is $ 2\sqrt{2} = 2.82$ \cite{katiyar2015investigation}.
The experimental value of maximum violation for $ \mathrm{\textbf{I}}_0, \mathrm{\textbf{I}}_1, \mathrm{\textbf{I}}_2,$ and $\mathrm{\textbf{I}}_3 $ are $ 2.40 \pm 0.017,~ 2.45 \pm 0.025,~ 2.39 \pm 0.016$, and $2.42 \pm 0.026$ respectively \cite{katiyar2015investigation}. There is a clear violation of the classical bound.  Reduced violation
than the theoretical value is due to $\mathrm{T}_2 $ decay and inhomogeneity in Radio Frequency (RF) pulses. 

\section{Ancilla Assisted Quantum State Tomography}
In experimental quantum information studies,
Quantum State Tomography (QST) is an important tool that is routinely used
to characterize an instantaneous quantum state \cite{chuangbook}.  
QST can be performed by a series of measurements of
noncommuting observables which together enables one to reconstruct 
the complete complex density matrix \cite{chuangbook}.  In the standard method,
the required number of independent experiments grows
exponentially with the number of input qubits \cite{ChuangPRSL98,ChuangPRA99}.
Anil Kumar and co-workers have illustrated QST using a single
two-dimensional NMR spectrum \cite{Aniltomo}. 
Later Nieuwenhuizen and co-workers showed how to reduce the number of independent experiments in the presence of an ancilla register \cite{Nieuwenhuizen}.
We referred to this method as Ancilla Assisted QST (AAQST) and
experimentally demonstrated it using NMR systems \cite{sutertomo,abhishekQST2013}.
AAQST also allows single shot mapping of density matrix which
not only reduces the experimental time, but also alleviates the need to 
prepare the target state repeatedly \cite{abhishekQST2013}.  

\begin{center}
\begin{figure}
\includegraphics[width=7cm]{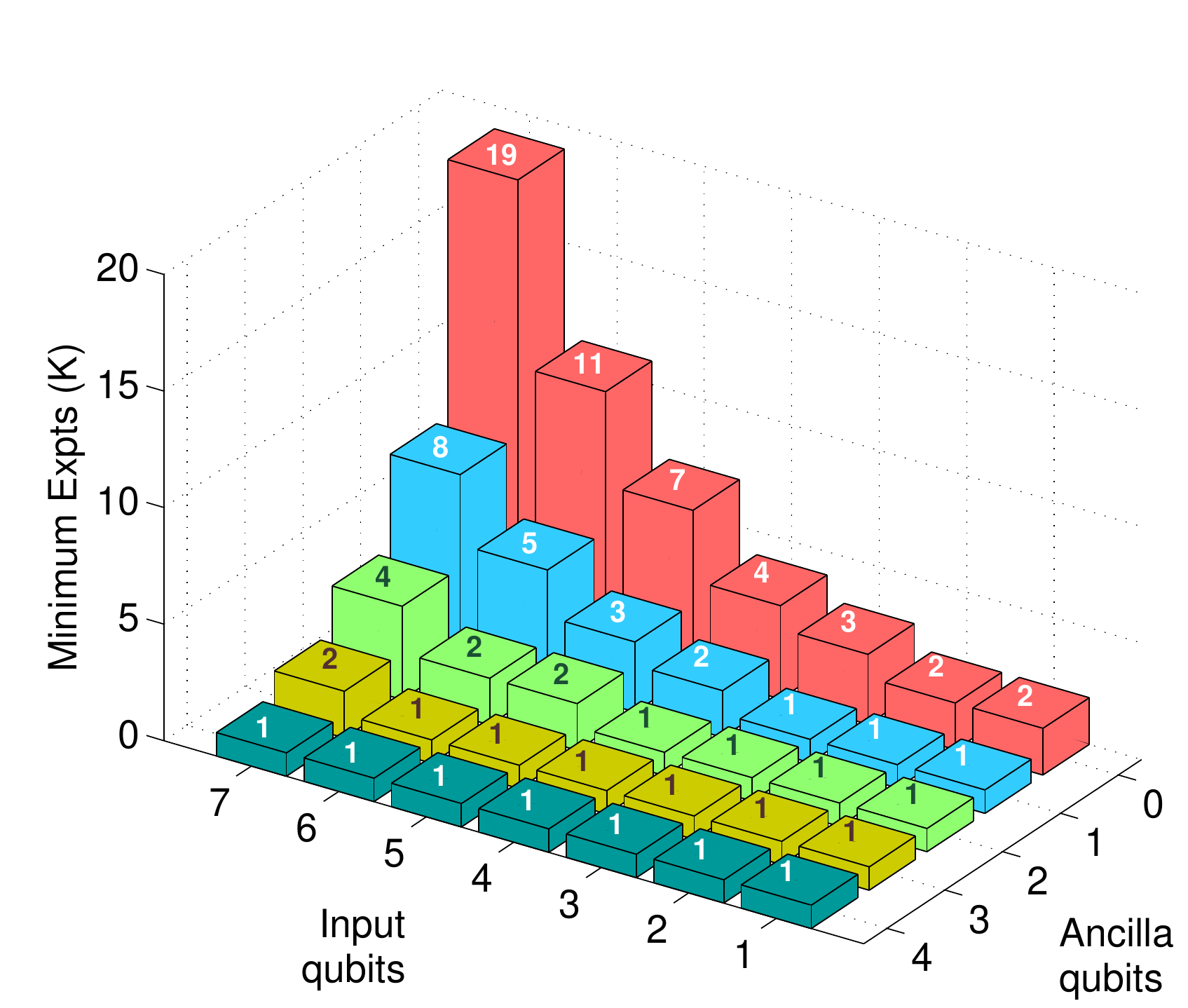} 
\caption{Minimum number of independent experiments required for QST
(with zero ancilla) and AAQST 
(Figure reproduced from reference \cite{abhishekQST2013}).  
}
\label{exptscaling} 
\end{figure}
\end{center}

To see how AAQST works, consider an input register of $n$-qubits associated with an ancilla register consisting of $\hat{n}$ qubits. 
The dimension of the combined system of $\tilde{n} = n+\hat{n}$ qubits 
is $\tilde{N} = N\hat{N}$, where $\hat{N} = 2^{\hat{n}}$.
A completely resolved NMR spectrum yields $\tilde{n}\tilde{N}$ real parameters.
We assume that the ancilla register begins with the maximally mixed initial state, with no contribution to the spectral lines from it.  
The deviation density matrix of the combined system is $\tilde{\rho} = \rho \otimes \mathbbm{1}/\hat{N}$. To perform AAQST, we apply a non-local unitary of the form,
\begin{eqnarray}
\tilde{U}_k = V \sum\limits_{a=0}^{\hat{N}-1} U_{ka} \otimes \vert a \rangle \langle a \vert.
\end{eqnarray}
Here $U_{ka}$ is the $k$th unitary on the input register dependent 
on the ancilla state $\vert a \rangle$ and $V$ is the local 
unitary on the ancilla.
The combined state evolves to
\begin{eqnarray}
&& \tilde{\rho}^{(k)} = \tilde{U}_k \tilde{\rho} \tilde{U}_k^\dagger \nonumber \\
&& = \frac{1}{\hat{N}}\sum\limits_{m,m',a} \rho_{mm'} U_{ka} \vert m \rangle \langle m' \vert U_{ka}^\dagger
\otimes V \vert a \rangle \langle a \vert V^\dagger.
\end{eqnarray}
Intensity of NMR spectrum is proportional to the observable
$\sum\limits_{j=1}^{\tilde{n}} \sigma_{jx}+i\sigma_{jy}$.
The spectrum of the combined system yields $\tilde{n}\tilde{N}$ linear
equations.  The minimum number of independent experiments
needed is now $O(N^2/(\tilde{n}\tilde{N}))$.
Choosing $\tilde{N} \gg N$, AAQST needs 
fewer than O($N/n$) experiments required in the standard QST.
In particular, when $\tilde{n}\tilde{N} \ge N^2$, a single
optimized unitary suffices for QST.
Fig. \ref{exptscaling} illustrates the minimum number ($K$) 
of experiments required for various sizes of input and ancilla 
registers. As illustrated, QST can be achieved with only one
experiment, if an ancilla of sufficient size is provided along with
\cite{abhishekQST2013}.

\begin{center}
\begin{figure}
\includegraphics[trim=4cm 0cm 5cm 0cm,clip=true,width=8.5cm]{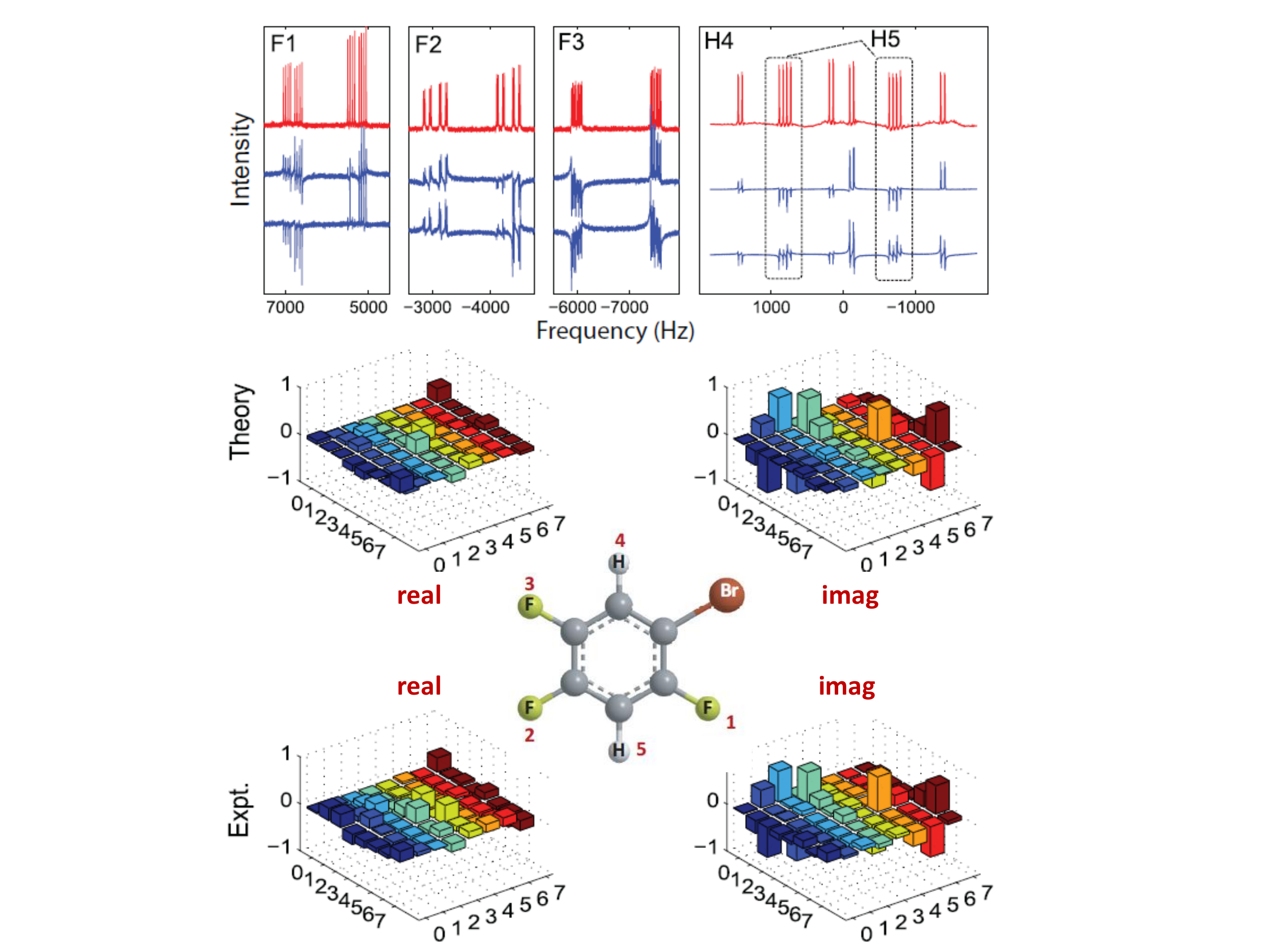} 
\caption{AAQST results for the state described in the text  \cite{abhishekQST2013}.
The reference spectrum is in the top trace.  The
real (middle trace) and the imaginary spectra (bottom trace) 
are obtained in a single shot AAQST experiment.
The bar plots correspond to theoretically expected
density matrices (top row) and those obtained from AAQST 
experiments (bottom row). 
The molecular structure and the labelling scheme of BTFBz 
is shown in the center
(Figure reproduced from reference \cite{abhishekQST2013}).
}
\label{aaqstres} 
\end{figure}
\end{center}

To demonstrate this procedure experimentally, we used
three $^{19}$F nuclei and two $^1$H nuclei
of 1-bromo-2,4,5-trifluorobenzene (BTFBz) partially oriented
in a liquid crystal namely, N-(4-methoxybenzaldehyde)-4-
butylaniline (MBBA) (Fig. \ref{aaqstres}) \cite{abhishekQST2013}. We chose 
the three $^{19}$F nuclei forming the input register
and two $^1$H nuclei forming the ancilla register.

Fig \ref{aaqstres} shows the experimental results corresponding
to a particular density matrix obtained by applying unitary
$U_0 = \left(\frac{\pi}{2}\right)_x^{F} \tau_0 (\pi)_x^{H} \tau_0 \left(\frac{\pi}{2}\right)_y^{F_1},$ with $\tau_0 = 2.5$ ms, on thermal equilibrium state.
The real and imaginary parts of the reconstructed density matrix 
along with the theoretically expected matrices are shown below the spectra 
in Fig. \ref{aaqstres}. 
Fidelity of the experimental state with the theoretical state was 0.95.
The entire three-qubit density matrix with 63 unknowns was
estimated by a single NMR experiment \cite{abhishekQST2013}.

\section{Single Scan Process Tomography}
Often one needs to characterize
the overall process acting on a quantum system.  Such a
characterization, achieved by a procedure called quantum
process tomography (QPT), is crucial in designing
fault-tolerant quantum processors  \cite{chuang97,zollar97}. 
QPT is realized by considering the quantum process as a map from
a complete set of initial states to final states, and experimentally
characterizing each of the final states using QST \cite{ChuangPRSL98}.
Since QST by itself involves repeated preparations of a target state,
QPT in general requires a number of independent experiments.
Therefore the total number of independent measurements required for QPT increases 
exponentially with the size of the system undergoing the process.

The physical realization of QPT has been demonstrated on various 
experimental setups \cite{ChuangPRA2001,QPTofQFT,EAPT1Exp,altepeter,obrian,bellstatefilter,qptITprl2006,QPT_IonTrapNature2010,QPT_SQUID,SQUID2009,MartiniSQUID2010,QPT2spSQUID,QPT_SQUIDChow2011,SQUID_Dewes2012,suterprotectedgate}.  
Several developments in the methodology of QPT have also been
reported \cite{simplifiedQPT1,simplifiedQPT2}.  In particular, it has been shown that ancilla assisted 
process tomography (AAPT) can characterize a process with a single QST
\cite{mazzei2003pauli,EAPT1Exp,PRLAriano2003,altepeter}.
By combining AAQST and AAPT, we showed that entire QPT can be
carried out with a single ensemble measurement  \cite{shukla2014single}.
We referred to this procedure as `single-scan quantum process tomography' (SSPT)
 \cite{shukla2014single}.

In the normal QPT procedure, the outcome of the process $\varepsilon$ is
expanded in a complete basis of
linearly independent elements $\{\rho_1, \rho_2, \cdots, \rho_{N^2}\}$ 
as well as using operator-sum representation, i.e.,
\begin{eqnarray}
\varepsilon(\rho_j) = \sum_k \lambda_{jk}\rho_k = \sum_i E_i \rho E_i^\dagger.
\label{lb}
\end{eqnarray}
The complex coefficients $\lambda_{jk}$ can be extracted using QST.
We can utilize a fixed set of basis operators $\{\tilde{E}_m\}$, and express 
$E_i = \sum_m e_{im} \tilde{E}_m$ so that
\begin{eqnarray}
\varepsilon(\rho) = \sum_{mn}\tilde{E}_m \rho \tilde{E}_n^\dagger \chi_{mn},
\label{tildeE}
\end{eqnarray}
where $\chi_{mn} = \sum_i e_{im} e_{in}^*$.  The $\chi$ matrix completely characterizes the process $\varepsilon$.
Since the set $\{\rho_k\}$ forms a complete basis, it is also possible to express
\begin{eqnarray}
\tilde{E}_m\rho_j\tilde{E}_n^\dagger = \sum_k \beta_{jk}^{mn}\rho_k,
\label{beta}
\end{eqnarray}
where $\beta_{jk}^{mn}$ can be calculated theoretically. Using eqns. \ref{lb}-\ref{beta} and using the linear independence of $\{\rho_k\}$,
we obtain
\begin{eqnarray}
\beta \chi = \lambda,
\label{chi}
\end{eqnarray}
from which $\chi$-matrix can be extracted by standard methods in linear algebra.

\begin{center}
\begin{figure}
\includegraphics[trim=0cm 0cm 0cm 0cm, clip=true,width=9.4cm]{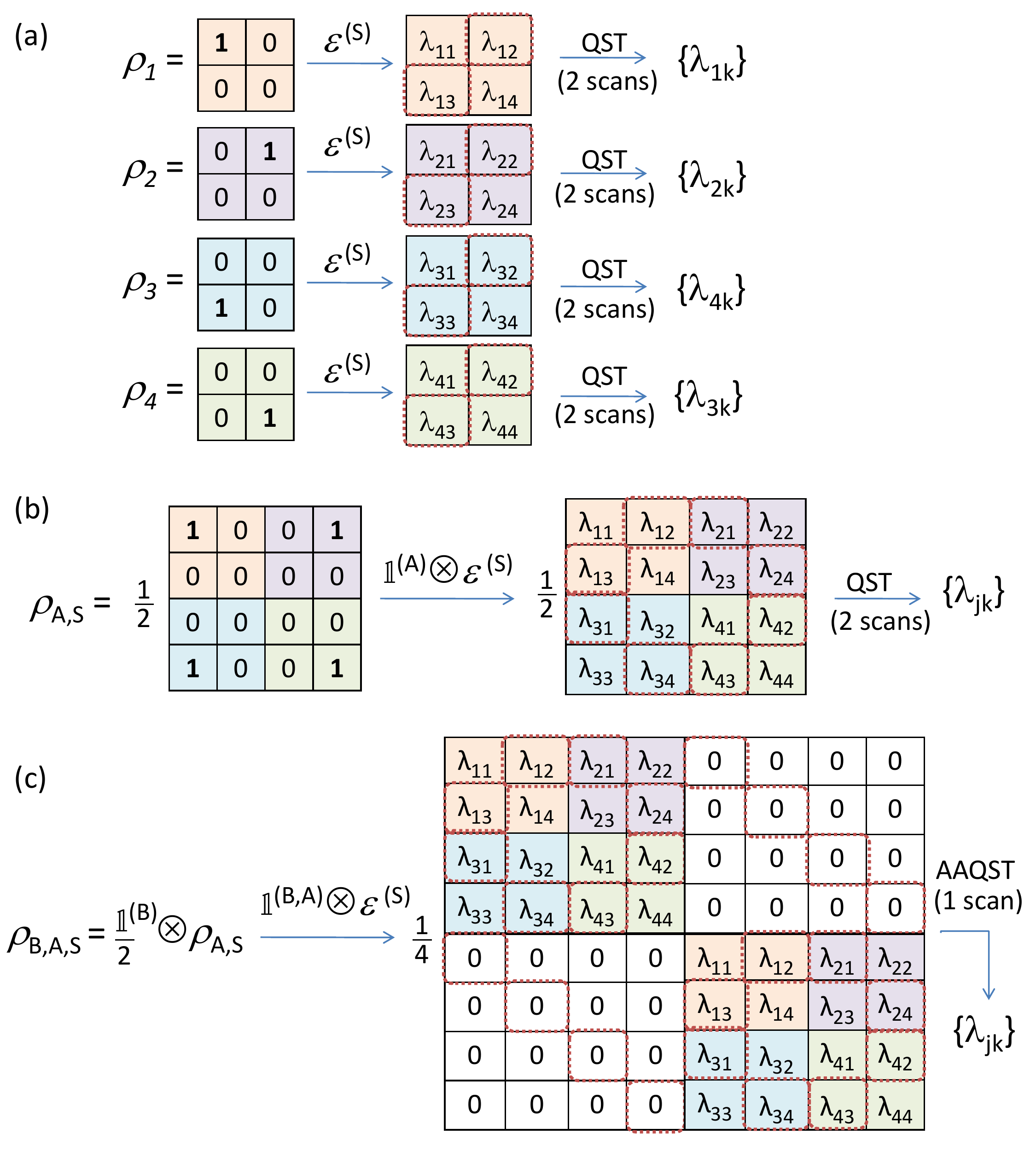} 
\caption{Illustrating (a) single-qubit QPT 
requiring a total of 8 NMR scans, (b) AAPT requiring 2 NMR scans, and (c) SSPT
requiring a single-scan NMR experiment \cite{shukla2014single}. In each case, dotted lines are used
to indicate the single-quantum elements of the density matrix which are directly observable.
Other elements are observed by converting them to observable single-quantum
coherences by using certain unitary operations in a subsequent scan(s)
(Figure reproduced from reference \cite{shukla2014single}).
}
\label{sum} 
\end{figure}
\end{center}

A comparison of QPT, AAPT, and SSPT procedures for a single qubit process is presented in Fig. \ref{sum} \cite{shukla2014single}.

Estimates of number of measurements for a small number of qubits
shown in the first column of Table 1 illustrate the exponential increase of $M_\mathrm{QPT}$ with $n$.
\begin{table}[h]
$
\begin{array}{|c|c|cc|cc|}
           \hline
           n & M_\mathrm{QPT} & M_\mathrm{AAPT} & (n_{A})& M_\mathrm{SSPT} &(n_{A}, n_{B})\\
           \hline
           \hline
           1      &     8     &      2 & (1)     &     1 &(1,1) \\
           \hline
           2    &      32      &     4 &(2)     &     1 &(2,2) \\
           \hline
           3    &     192      &    11 &(3)     &     1 &(3,3) \\
           \hline
           4    &    1024      &    32 &(4)     &     1 &(4,5) \\
           \hline
           5   &     7168    &     103 &(5)     &     1 &(5,6) \\
           \hline
\end{array}
$
\caption{Comparison of number of scans and 
number of ancilla qubits (in parenthesis) required for 
$n$-qubit QPT, AAPT, and SSPT.}
\end{table}

\begin{figure}
\includegraphics[trim=4cm 0cm 4cm 0cm, clip=true,width=8.8cm]{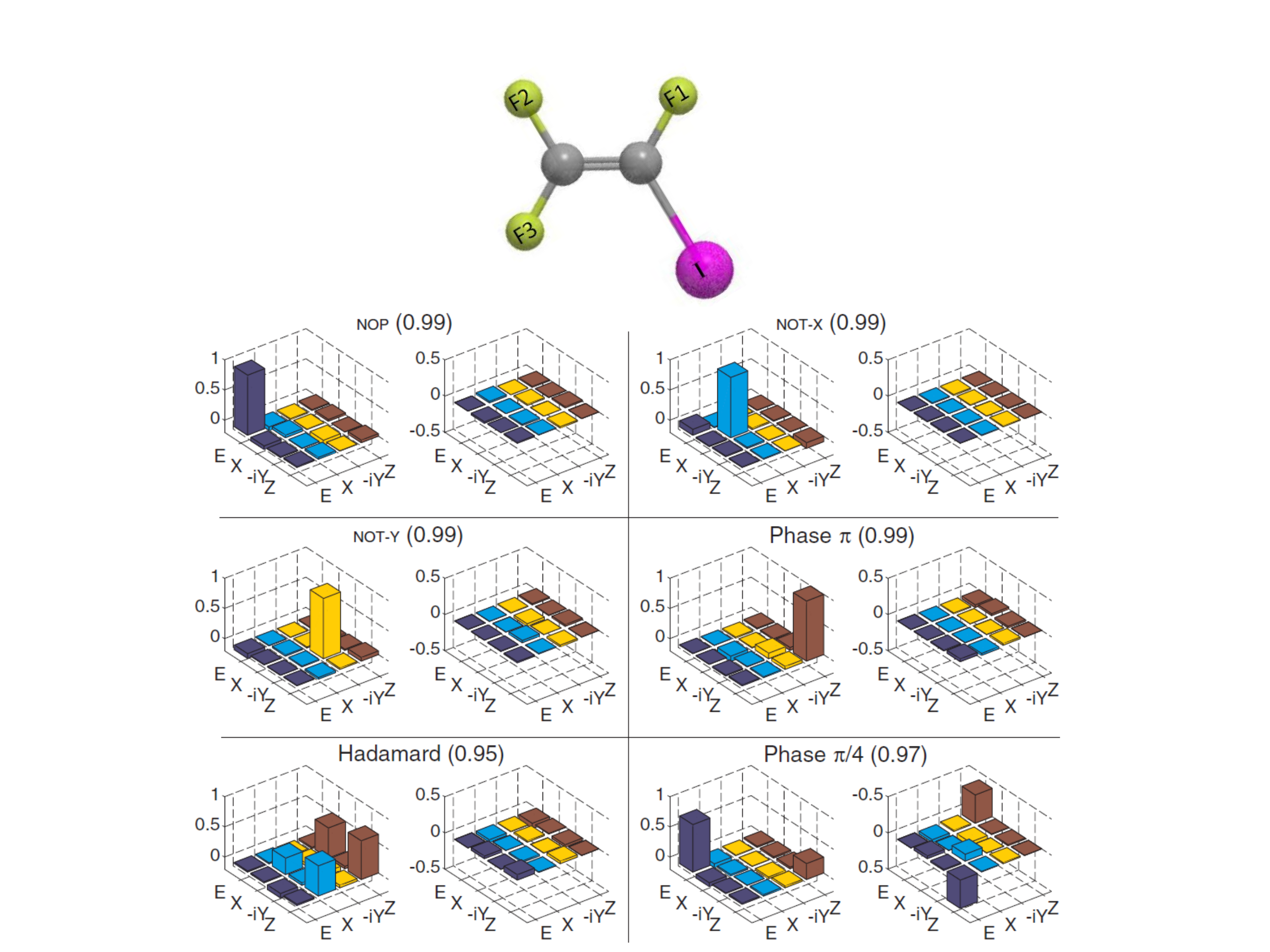} 
\caption{The barplots showing experimental $\chi$-matrices 
for various quantum processes obtained using SSPT.  In each case, the left and right barplots correspond to the real and imaginary parts respectively, and the fidelities are indicated in parenthesis
(Figure reproduced from reference \cite{shukla2014single}).
} 
\label{ssptres}
\end{figure}

The experimental demonstration of a single-qubit SSPT was carried out using iodotrifluoroethylene dissolved in acetone-D$_6$ as a 3-qubit system Fig. \ref{ssptres} \cite{shukla2014single}.  The experimentally obtained $\chi$ matrices for certain quantum processes using the single scan procedure are shown in Fig. \ref{ssptres}.

\section{Ancilla assisted noise engineering}
Preserving coherence is a very important aspect to realize quantum processors, and 
hence various techniques have been developed to suppress decoherence. They include dynamical decoupling (DD) techniques \cite{cp,lloyd1,lloyd2,uhrig}, quantum error correction \cite{preskill}, adiabatic quantum computation \cite{farhi}, and use of decoherence-free subspaces \cite{DFS}. 
Earlier Teklemariam et al. introduced artificial decoherence by
achieving irreversible phase damping via constant perturbation of the environment qubits (by random classical fields), and thus mimicking a large dimensional environmental bath. Such experiments provide insights about decoherence processes and may pave the way for improving decoherence suppression techniques.

In our work we simulated such a decoherence process on a NMR simulator with two qubits, where one acts as the system and the other as environment
\cite{hegde2014engineered}. We then subjected the system qubit to certain DD 
sequences and observed their competition with the engineered decoherence
through noise spectroscopy.

The two qubit register was initially in the product state,
$\rho(0) = \rho^s(0) \otimes \rho^e(0)$.
Here $\rho^s(0)$ is the system state and $\rho^e(0)$ is the environment state. 
We chose $^1$H and $^{13}$C nuclear spins in $^{13}$C-labelled chloroform ($^{13}$CHCl$_3$ 
dissolved in CDCl$_3$) as the system and environment qubit respectively.  
The NMR Hamiltonian is
\begin{equation}
{\cal H}=\pi(\nu_s\sigma_z^{s}+\nu_e\sigma_z^{e}+\frac{J}{2}\sigma_z^{s}\sigma_z^{e}),
\label{Ham}
\end{equation}
where $\nu_s$ and $\nu_e$ are the resonant frequencies of the system and the environment qubits respectively, $J$ is the coupling strength between the two, and $\sigma_z^{s}$, $\sigma_z^{e}$ are the Pauli operators. 
In a total duration $T$, the propagator $U = e^{-i{\cal H}T}$
entangles the system qubit with the environment qubit via the interaction $J$.
We engineered decoherence by a series of RF kicks of arbitrary angles $\epsilon \in [-\theta,\theta]$ on the environment qubit. These kicks induced artificial decoherence on the system qubit. 
Teklemariam et al. proved that induced decoherence of the system qubit
depends on the kick-rate $\Gamma$, range of kick-angles $\theta$, and coupling strength $J$ \cite{cory}. Their model predicted that for small kick-angles $\epsilon$ and for lower kick rates $\Gamma$, 
decoherence rate $1/T_2$ increases linearly with $\Gamma$. After a certain value of  $\Gamma$,
$1/T_2$ saturates, and then onwards, it decreases exponentially with $\Gamma$.  

\begin{figure}
\centering
\includegraphics[width=8cm]{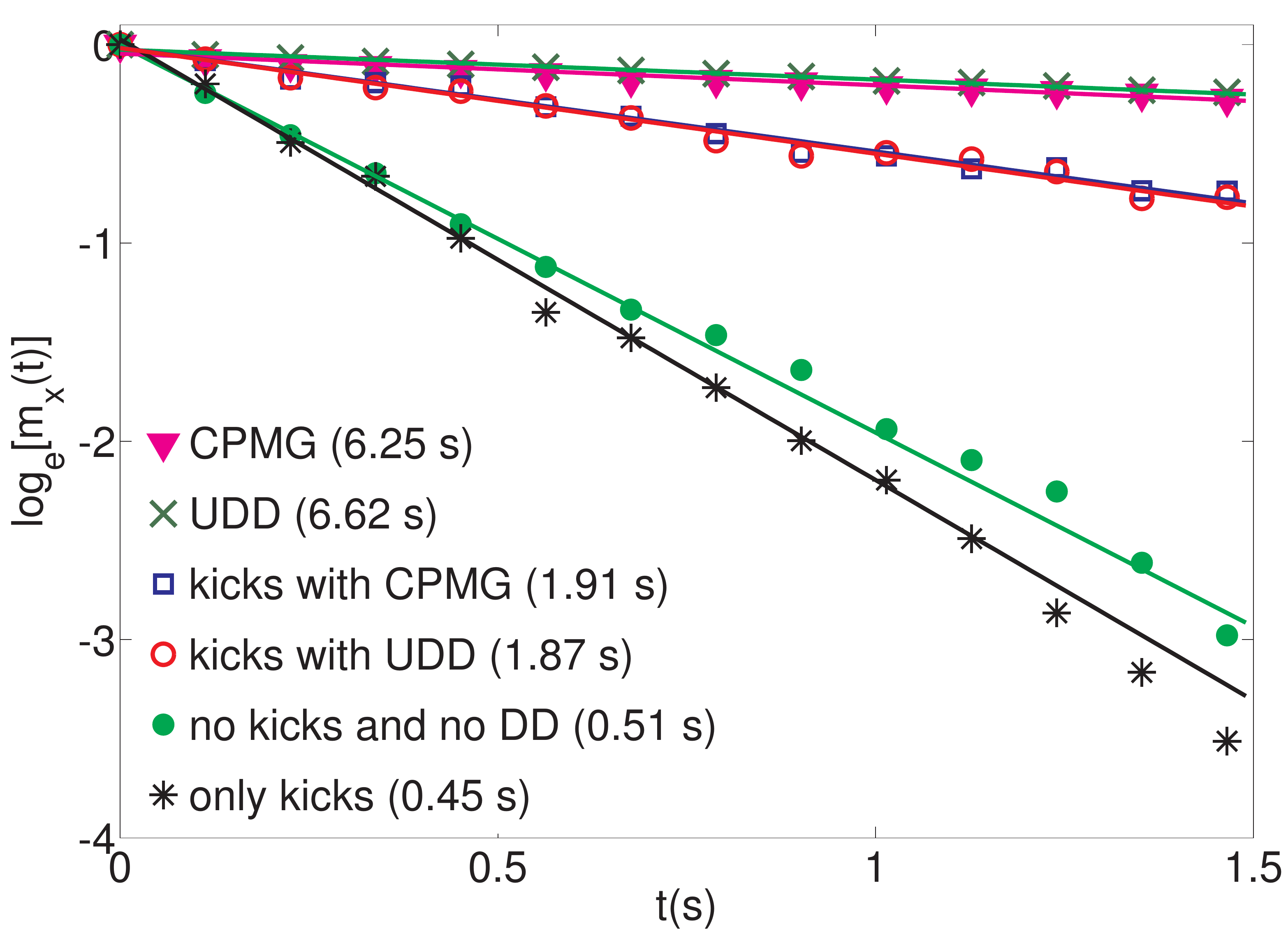}
\caption{Logarithm of transverse magnetization $\log(M_x)$ as a function of time under different 
pulse sequences as indicated in the legend. Here $\tau = 3.2$ ms,
$\Gamma = 25$ kicks/ms, and $\epsilon \in [0^\circ,1^\circ]$. The $T_2$ values for various pulse sequences are shown in the legend
(Figure reproduced from reference \cite{hegde2014engineered}).
}
\label{comp_m}
\end{figure}

Our experimental results for $\epsilon \in [0^\circ,1^\circ]$ and $\Gamma = 25$ kicks/ms
are shown in Fig. \ref{comp_m} (indicated by stars).  
For comparison we have also shown the decay of magnetization without kicks (indicated by filled circles).  
Evidently, the kicks on environment have introduced additional decoherence thereby increasing the decay of system coherence \cite{hegde2014engineered}.

Dynamical decoupling attempts to inhibit decoherence of system by 
rapid modulation of the system state so as to cancel the system-environment
joint evolutions.  The two standard 
DD sequences are: (i) CPMG \cite{cp} and (ii) UDD \cite{uhrig}.
CPMG consists of a series of equidistant $\pi$ pulses applied in the system qubits. 
In an N-pulse UDD of cycle time $t_c$,  the time instant $t_j$ of $j^\mathrm{th}$ $\pi$-pulse is given by 
$t_j = t_c \sin^2 \left[\frac{\pi j}{2(N+1)}\right]$.
The results of the experiments for $t_c = 22.4$ ms and with different
kick-parameters are shown in Fig. \ref{comp_m}.  The competition between
kicks-induced decoherence and DD sequences can be readily observed \cite{hegde2014engineered}.

Noise spectroscopy provides information about noise spectral density, which is the frequency distribution of noise and is helpful in not only understanding the performance of standard DD sequences, but also in optimizing them \cite{biercuk,biercuk1,pan}. 
In the limit of a large number of $\pi$ pulses, the CPMG filter function resembles a delta peak at $\omega$, and samples this particular spectral frequency. The amplitude of the noise $S(\omega)$ can be determined by 
using the relation \cite{yuge} $S(\omega) \simeq \frac{\pi^2}{4 T_2(\omega)}$.
Thus by measuring $T_2(\omega)$ for a range of $\omega=\pi/\tau$ values, we can scan the profile of $S(\omega)$. 

\begin{figure}[t]
\centering
\includegraphics[width=8.7cm]{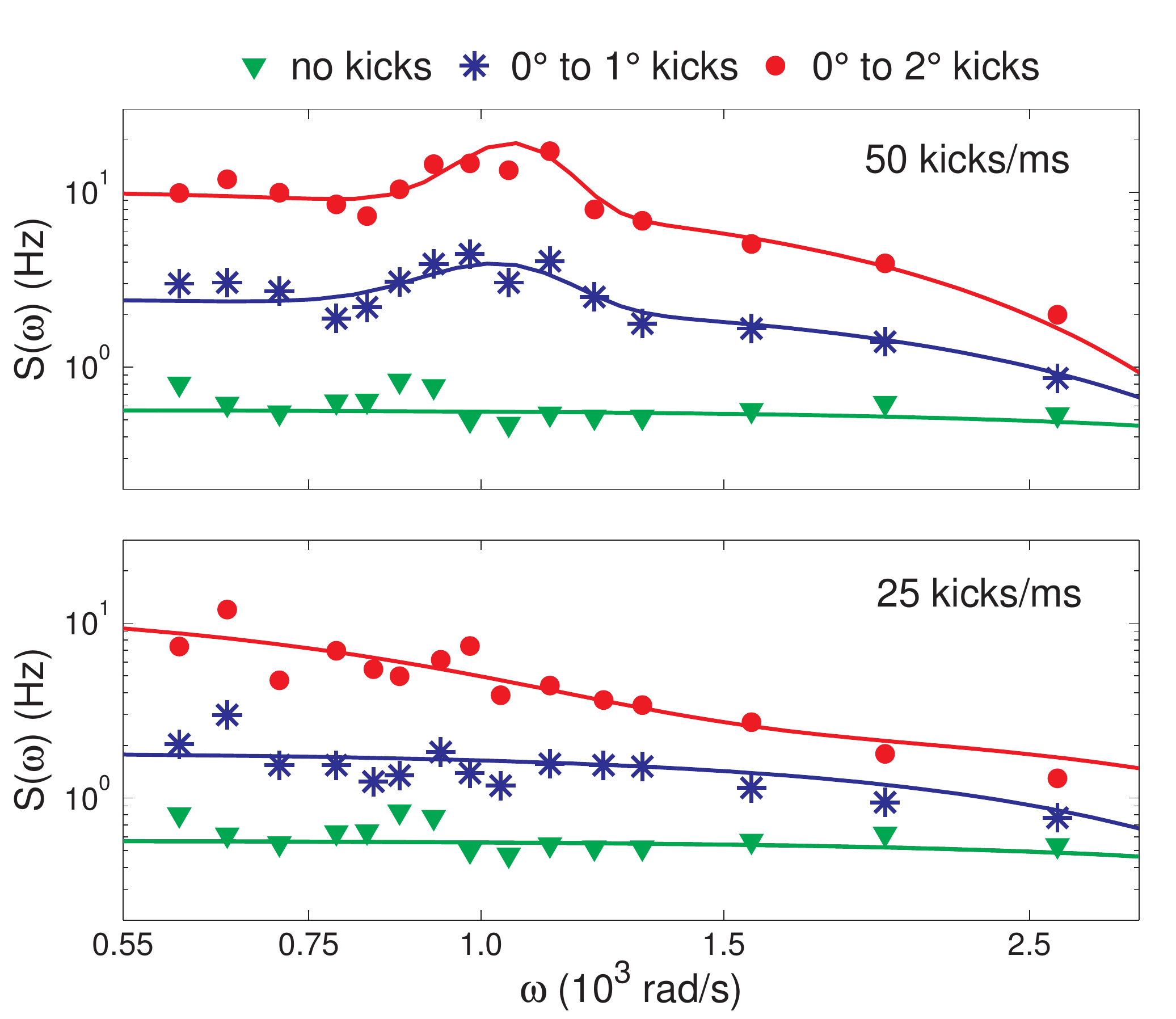}
\caption{Experimental spectral density profiles with different kick-angles
(as indicated in the legend) and with kick-rates 50 kicks/ms (top trace)
and 25 kicks/ms (bottom trace). In both the traces, experimental spectral
density profile without kicks is also shown for comparison. The smooth
lines correspond to fits with one or two Gaussians
(Figure reproduced from reference \cite{hegde2014engineered}).}
\label{sd}
\end{figure}

The experimental spectral density profiles
of only natural decoherence (lowest curve in each sub-plot), and with kicks of
different kick-parameters are shown in Fig. \ref{sd}  \cite{hegde2014engineered}.
Clearly the effect of kicks is to increase the area under the spectral
density profiles and thereby leading to faster decoherence.  Moreover, 
for a given kick-rate $\Gamma$, larger the range of kick-angles, higher is the 
spectral density profile \cite{hegde2014engineered}.  

\section{Summary}
In this review article, we described several recent protocols for efficient 
measurements on quantum systems, and illustrated their NMR implementations.

In section II, we described ancilla-assisted noninvasive measurements, where
the measurement result of an intermediate observable was temporarily stored in
an ancilla qubit.  A final joint-measurement of the system-ancilla register
revealed the joint probabilities in a noninvasive way.  We also showed
the application of this technique in studying entropic Leggett-Garg inequality
\cite{elgimahesh}.

In section III, we described extracting expectation values of various types
of operators.  Applications of these methods are illustrated in the estimation of Franck-Condon coefficients and in the investigation of quantum contextuality
\cite{joshi2014estimating,katiyar2015investigation}.

In section IV and V we described efficient ways to characterize quantum states and quantum process by exploiting ancilla qubits.  We also illustrated single-scan quantum state tomography as well as single-scan quantum process tomography using NMR systems.  These techniques not only alleviate the need of repeated measurements, but also allow the study of random states or dynamic processes
\cite{abhishekQST2013,shukla2014single}.  

Finally in section VI we described
ancilla-assisted noise engineering, where random fields applied to the ancilla
qubits cause controllable decoherence on the system qubits.  We illustrated this
phenomena using a two-qubit NMR system, and studied the engineered decoherence by
measuring noise spectrum \cite{hegde2014engineered}.

Although we have used NMR techniques to demonstrate the above protocols experimentally, these procedures are quite general in nature, and can easily be adopted to other experimental techniques as well. 

\section{Acknowledgements}
The authors are grateful to Prof. Anil Kumar of IISc Bangalore, 
Prof. Usha Devi of Bangalore University, Prof. A. K. Rajagopal of Inspire Institute, USA, and Dr. Anirban Hazra of IISER, Pune for discussions. Projects described in this article were partly supported by DST Project No. SR/S2/LOP-0017/2009.

\bibliographystyle{apsrev4-1}
\bibliography{biblio1}

\begin{thebibliography}{66}%
\makeatletter
\providecommand \@ifxundefined [1]{%
 \@ifx{#1\undefined}
}%
\providecommand \@ifnum [1]{%
 \ifnum #1\expandafter \@firstoftwo
 \else \expandafter \@secondoftwo
 \fi
}%
\providecommand \@ifx [1]{%
 \ifx #1\expandafter \@firstoftwo
 \else \expandafter \@secondoftwo
 \fi
}%
\providecommand \natexlab [1]{#1}%
\providecommand \enquote  [1]{``#1''}%
\providecommand \bibnamefont  [1]{#1}%
\providecommand \bibfnamefont [1]{#1}%
\providecommand \citenamefont [1]{#1}%
\providecommand \href@noop [0]{\@secondoftwo}%
\providecommand \href [0]{\begingroup \@sanitize@url \@href}%
\providecommand \@href[1]{\@@startlink{#1}\@@href}%
\providecommand \@@href[1]{\endgroup#1\@@endlink}%
\providecommand \@sanitize@url [0]{\catcode `\\12\catcode `\$12\catcode
  `\&12\catcode `\#12\catcode `\^12\catcode `\_12\catcode `\%12\relax}%
\providecommand \@@startlink[1]{}%
\providecommand \@@endlink[0]{}%
\providecommand \url  [0]{\begingroup\@sanitize@url \@url }%
\providecommand \@url [1]{\endgroup\@href {#1}{\urlprefix }}%
\providecommand \urlprefix  [0]{URL }%
\providecommand \Eprint [0]{\href }%
\providecommand \doibase [0]{http://dx.doi.org/}%
\providecommand \selectlanguage [0]{\@gobble}%
\providecommand \bibinfo  [0]{\@secondoftwo}%
\providecommand \bibfield  [0]{\@secondoftwo}%
\providecommand \translation [1]{[#1]}%
\providecommand \BibitemOpen [0]{}%
\providecommand \bibitemStop [0]{}%
\providecommand \bibitemNoStop [0]{.\EOS\space}%
\providecommand \EOS [0]{\spacefactor3000\relax}%
\providecommand \BibitemShut  [1]{\csname bibitem#1\endcsname}%
\let\auto@bib@innerbib\@empty
\bibitem [{\citenamefont {Athalye}\ \emph
  {et~al.}(2011{\natexlab{a}})\citenamefont {Athalye}, \citenamefont {Roy},\
  and\ \citenamefont {Mahesh}}]{elgimahesh}%
  \BibitemOpen
  \bibfield  {author} {\bibinfo {author} {\bibfnamefont {V.}~\bibnamefont
  {Athalye}}, \bibinfo {author} {\bibfnamefont {S.~S.}\ \bibnamefont {Roy}}, \
  and\ \bibinfo {author} {\bibfnamefont {T.~S.}\ \bibnamefont {Mahesh}},\
  }\href {\doibase 10.1103/PhysRevLett.107.130402} {\bibfield  {journal}
  {\bibinfo  {journal} {Phys. Rev. Lett.}\ }\textbf {\bibinfo {volume} {107}},\
  \bibinfo {pages} {130402} (\bibinfo {year} {2011}{\natexlab{a}})}\BibitemShut
  {NoStop}%
\bibitem [{\citenamefont {Knee}\ \emph
  {et~al.}(2012{\natexlab{a}})\citenamefont {Knee}, \citenamefont {Gauger},
  \citenamefont {Briggs},\ and\ \citenamefont {Benjamin}}]{knee}%
  \BibitemOpen
  \bibfield  {author} {\bibinfo {author} {\bibfnamefont {G.~C.}\ \bibnamefont
  {Knee}}, \bibinfo {author} {\bibfnamefont {E.~M.}\ \bibnamefont {Gauger}},
  \bibinfo {author} {\bibfnamefont {G.~A.~D.}\ \bibnamefont {Briggs}}, \ and\
  \bibinfo {author} {\bibfnamefont {S.~C.}\ \bibnamefont {Benjamin}},\ }\href
  {http://stacks.iop.org/1367-2630/14/i=5/a=058001} {\bibfield  {journal}
  {\bibinfo  {journal} {New Journal of Physics}\ }\textbf {\bibinfo {volume}
  {14}},\ \bibinfo {pages} {058001} (\bibinfo {year}
  {2012}{\natexlab{a}})}\BibitemShut {NoStop}%
\bibitem [{\citenamefont {Leggett}\ and\ \citenamefont {Garg}(1985)}]{lgi1985}%
  \BibitemOpen
  \bibfield  {author} {\bibinfo {author} {\bibfnamefont {A.~J.}\ \bibnamefont
  {Leggett}}\ and\ \bibinfo {author} {\bibfnamefont {A.}~\bibnamefont {Garg}},\
  }\href {\doibase 10.1103/PhysRevLett.54.857} {\bibfield  {journal} {\bibinfo
  {journal} {Phys. Rev. Lett.}\ }\textbf {\bibinfo {volume} {54}},\ \bibinfo
  {pages} {857} (\bibinfo {year} {1985})}\BibitemShut {NoStop}%
\bibitem [{\citenamefont {Palacios-Laloy}\ \emph {et~al.}(2010)\citenamefont
  {Palacios-Laloy}, \citenamefont {Mallet}, \citenamefont {Nguyen},
  \citenamefont {Bertet}, \citenamefont {Vion}, \citenamefont {Esteve},\ and\
  \citenamefont {Korotkov}}]{palacios2010experimental}%
  \BibitemOpen
  \bibfield  {author} {\bibinfo {author} {\bibfnamefont {A.}~\bibnamefont
  {Palacios-Laloy}}, \bibinfo {author} {\bibfnamefont {F.}~\bibnamefont
  {Mallet}}, \bibinfo {author} {\bibfnamefont {F.}~\bibnamefont {Nguyen}},
  \bibinfo {author} {\bibfnamefont {P.}~\bibnamefont {Bertet}}, \bibinfo
  {author} {\bibfnamefont {D.}~\bibnamefont {Vion}}, \bibinfo {author}
  {\bibfnamefont {D.}~\bibnamefont {Esteve}}, \ and\ \bibinfo {author}
  {\bibfnamefont {A.~N.}\ \bibnamefont {Korotkov}},\ }\href@noop {} {\bibfield
  {journal} {\bibinfo  {journal} {Nature Physics}\ }\textbf {\bibinfo {volume}
  {6}},\ \bibinfo {pages} {442} (\bibinfo {year} {2010})}\BibitemShut {NoStop}%
\bibitem [{\citenamefont {Athalye}\ \emph
  {et~al.}(2011{\natexlab{b}})\citenamefont {Athalye}, \citenamefont {Roy},\
  and\ \citenamefont {Mahesh}}]{athalye2011investigation}%
  \BibitemOpen
  \bibfield  {author} {\bibinfo {author} {\bibfnamefont {V.}~\bibnamefont
  {Athalye}}, \bibinfo {author} {\bibfnamefont {S.~S.}\ \bibnamefont {Roy}}, \
  and\ \bibinfo {author} {\bibfnamefont {T.~S.}\ \bibnamefont {Mahesh}},\
  }\href@noop {} {\bibfield  {journal} {\bibinfo  {journal} {Physical review
  letters}\ }\textbf {\bibinfo {volume} {107}},\ \bibinfo {pages} {130402}
  (\bibinfo {year} {2011}{\natexlab{b}})}\BibitemShut {NoStop}%
\bibitem [{\citenamefont {Lambert}\ \emph {et~al.}(2011)\citenamefont
  {Lambert}, \citenamefont {Johansson},\ and\ \citenamefont
  {Nori}}]{Lambert2011}%
  \BibitemOpen
  \bibfield  {author} {\bibinfo {author} {\bibfnamefont {N.}~\bibnamefont
  {Lambert}}, \bibinfo {author} {\bibfnamefont {R.}~\bibnamefont {Johansson}},
  \ and\ \bibinfo {author} {\bibfnamefont {F.}~\bibnamefont {Nori}},\ }\href
  {\doibase 10.1103/PhysRevB.84.245421} {\bibfield  {journal} {\bibinfo
  {journal} {Phys. Rev. B}\ }\textbf {\bibinfo {volume} {84}},\ \bibinfo
  {pages} {245421} (\bibinfo {year} {2011})}\BibitemShut {NoStop}%
\bibitem [{\citenamefont {Goggin}\ \emph {et~al.}(2011)\citenamefont {Goggin},
  \citenamefont {Almeida}, \citenamefont {Barbieri}, \citenamefont {Lanyon},
  \citenamefont {O’Brien}, \citenamefont {White},\ and\ \citenamefont
  {Pryde}}]{goggin2011violation}%
  \BibitemOpen
  \bibfield  {author} {\bibinfo {author} {\bibfnamefont {M.}~\bibnamefont
  {Goggin}}, \bibinfo {author} {\bibfnamefont {M.}~\bibnamefont {Almeida}},
  \bibinfo {author} {\bibfnamefont {M.}~\bibnamefont {Barbieri}}, \bibinfo
  {author} {\bibfnamefont {B.}~\bibnamefont {Lanyon}}, \bibinfo {author}
  {\bibfnamefont {J.}~\bibnamefont {O’Brien}}, \bibinfo {author}
  {\bibfnamefont {A.}~\bibnamefont {White}}, \ and\ \bibinfo {author}
  {\bibfnamefont {G.}~\bibnamefont {Pryde}},\ }\href@noop {} {\bibfield
  {journal} {\bibinfo  {journal} {Proceedings of the National Academy of
  Sciences}\ }\textbf {\bibinfo {volume} {108}},\ \bibinfo {pages} {1256}
  (\bibinfo {year} {2011})}\BibitemShut {NoStop}%
\bibitem [{\citenamefont {Dressel}\ \emph {et~al.}(2011)\citenamefont
  {Dressel}, \citenamefont {Broadbent}, \citenamefont {Howell},\ and\
  \citenamefont {Jordan}}]{jordanPRL2011}%
  \BibitemOpen
  \bibfield  {author} {\bibinfo {author} {\bibfnamefont {J.}~\bibnamefont
  {Dressel}}, \bibinfo {author} {\bibfnamefont {C.~J.}\ \bibnamefont
  {Broadbent}}, \bibinfo {author} {\bibfnamefont {J.~C.}\ \bibnamefont
  {Howell}}, \ and\ \bibinfo {author} {\bibfnamefont {A.~N.}\ \bibnamefont
  {Jordan}},\ }\href {\doibase 10.1103/PhysRevLett.106.040402} {\bibfield
  {journal} {\bibinfo  {journal} {Phys. Rev. Lett.}\ }\textbf {\bibinfo
  {volume} {106}},\ \bibinfo {pages} {040402} (\bibinfo {year}
  {2011})}\BibitemShut {NoStop}%
\bibitem [{\citenamefont {Souza}\ \emph {et~al.}(2011)\citenamefont {Souza},
  \citenamefont {Oliveira},\ and\ \citenamefont
  {Sarthour}}]{Oliveira2011timeinequality}%
  \BibitemOpen
  \bibfield  {author} {\bibinfo {author} {\bibfnamefont {A.}~\bibnamefont
  {Souza}}, \bibinfo {author} {\bibfnamefont {I.}~\bibnamefont {Oliveira}}, \
  and\ \bibinfo {author} {\bibfnamefont {R.}~\bibnamefont {Sarthour}},\
  }\href@noop {} {\bibfield  {journal} {\bibinfo  {journal} {New Journal of
  Physics}\ }\textbf {\bibinfo {volume} {13}},\ \bibinfo {pages} {053023}
  (\bibinfo {year} {2011})}\BibitemShut {NoStop}%
\bibitem [{\citenamefont {Emary}(2012)}]{etrans}%
  \BibitemOpen
  \bibfield  {author} {\bibinfo {author} {\bibfnamefont {C.}~\bibnamefont
  {Emary}},\ }\href {\doibase 10.1103/PhysRevB.86.085418} {\bibfield  {journal}
  {\bibinfo  {journal} {Phys. Rev. B}\ }\textbf {\bibinfo {volume} {86}},\
  \bibinfo {pages} {085418} (\bibinfo {year} {2012})}\BibitemShut {NoStop}%
\bibitem [{\citenamefont {Suzuki}\ \emph {et~al.}(2012)\citenamefont {Suzuki},
  \citenamefont {Iinuma},\ and\ \citenamefont {Hofmann}}]{suzuki2012violation}%
  \BibitemOpen
  \bibfield  {author} {\bibinfo {author} {\bibfnamefont {Y.}~\bibnamefont
  {Suzuki}}, \bibinfo {author} {\bibfnamefont {M.}~\bibnamefont {Iinuma}}, \
  and\ \bibinfo {author} {\bibfnamefont {H.~F.}\ \bibnamefont {Hofmann}},\
  }\href@noop {} {\bibfield  {journal} {\bibinfo  {journal} {New Journal of
  Physics}\ }\textbf {\bibinfo {volume} {14}},\ \bibinfo {pages} {103022}
  (\bibinfo {year} {2012})}\BibitemShut {NoStop}%
\bibitem [{\citenamefont {Yong-Nan}\ \emph {et~al.}(2012)\citenamefont
  {Yong-Nan}, \citenamefont {Yang}, \citenamefont {Rong-Chun}, \citenamefont
  {Jian-Shun},\ and\ \citenamefont {Chuan-Feng}}]{violationQuantumdots}%
  \BibitemOpen
  \bibfield  {author} {\bibinfo {author} {\bibfnamefont {S.}~\bibnamefont
  {Yong-Nan}}, \bibinfo {author} {\bibfnamefont {Z.}~\bibnamefont {Yang}},
  \bibinfo {author} {\bibfnamefont {G.}~\bibnamefont {Rong-Chun}}, \bibinfo
  {author} {\bibfnamefont {T.}~\bibnamefont {Jian-Shun}}, \ and\ \bibinfo
  {author} {\bibfnamefont {L.}~\bibnamefont {Chuan-Feng}},\ }\href@noop {}
  {\bibfield  {journal} {\bibinfo  {journal} {Chinese Physics Letters}\
  }\textbf {\bibinfo {volume} {29}},\ \bibinfo {pages} {120302} (\bibinfo
  {year} {2012})}\BibitemShut {NoStop}%
\bibitem [{\citenamefont {Zhou}\ \emph {et~al.}(2012)\citenamefont {Zhou},
  \citenamefont {Huelga}, \citenamefont {Li},\ and\ \citenamefont
  {Guo}}]{zhou2012experimental}%
  \BibitemOpen
  \bibfield  {author} {\bibinfo {author} {\bibfnamefont {Z.-Q.}\ \bibnamefont
  {Zhou}}, \bibinfo {author} {\bibfnamefont {S.~F.}\ \bibnamefont {Huelga}},
  \bibinfo {author} {\bibfnamefont {C.-F.}\ \bibnamefont {Li}}, \ and\ \bibinfo
  {author} {\bibfnamefont {G.-C.}\ \bibnamefont {Guo}},\ }\href@noop {}
  {\bibfield  {journal} {\bibinfo  {journal} {arXiv preprint arXiv:1209.2176}\
  } (\bibinfo {year} {2012})}\BibitemShut {NoStop}%
\bibitem [{\citenamefont {Knee}\ \emph
  {et~al.}(2012{\natexlab{b}})\citenamefont {Knee}, \citenamefont {Simmons},
  \citenamefont {Gauger}, \citenamefont {Morton}, \citenamefont {Riemann},
  \citenamefont {Abrosimov}, \citenamefont {Becker}, \citenamefont {Pohl},
  \citenamefont {Itoh}, \citenamefont {Thewalt} \emph
  {et~al.}}]{knee2012violation}%
  \BibitemOpen
  \bibfield  {author} {\bibinfo {author} {\bibfnamefont {G.~C.}\ \bibnamefont
  {Knee}}, \bibinfo {author} {\bibfnamefont {S.}~\bibnamefont {Simmons}},
  \bibinfo {author} {\bibfnamefont {E.~M.}\ \bibnamefont {Gauger}}, \bibinfo
  {author} {\bibfnamefont {J.~J.}\ \bibnamefont {Morton}}, \bibinfo {author}
  {\bibfnamefont {H.}~\bibnamefont {Riemann}}, \bibinfo {author} {\bibfnamefont
  {N.~V.}\ \bibnamefont {Abrosimov}}, \bibinfo {author} {\bibfnamefont
  {P.}~\bibnamefont {Becker}}, \bibinfo {author} {\bibfnamefont {H.-J.}\
  \bibnamefont {Pohl}}, \bibinfo {author} {\bibfnamefont {K.~M.}\ \bibnamefont
  {Itoh}}, \bibinfo {author} {\bibfnamefont {M.~L.}\ \bibnamefont {Thewalt}},
  \emph {et~al.},\ }\href@noop {} {\bibfield  {journal} {\bibinfo  {journal}
  {Nature communications}\ }\textbf {\bibinfo {volume} {3}},\ \bibinfo {pages}
  {606} (\bibinfo {year} {2012}{\natexlab{b}})}\BibitemShut {NoStop}%
\bibitem [{\citenamefont {Devi}\ \emph {et~al.}(2013)\citenamefont {Devi},
  \citenamefont {Karthik}, \citenamefont {Sudha},\ and\ \citenamefont
  {Rajagopal}}]{elgiUshadevi}%
  \BibitemOpen
  \bibfield  {author} {\bibinfo {author} {\bibfnamefont {A.~R.~U.}\
  \bibnamefont {Devi}}, \bibinfo {author} {\bibfnamefont {H.~S.}\ \bibnamefont
  {Karthik}}, \bibinfo {author} {\bibnamefont {Sudha}}, \ and\ \bibinfo
  {author} {\bibfnamefont {A.~K.}\ \bibnamefont {Rajagopal}},\ }\href {\doibase
  10.1103/PhysRevA.87.052103} {\bibfield  {journal} {\bibinfo  {journal} {Phys.
  Rev. A}\ }\textbf {\bibinfo {volume} {87}},\ \bibinfo {pages} {052103}
  (\bibinfo {year} {2013})}\BibitemShut {NoStop}%
\bibitem [{\citenamefont {Karthik}\ \emph {et~al.}(2013)\citenamefont
  {Karthik}, \citenamefont {Katiyar}, \citenamefont {Shukla}, \citenamefont
  {Mahesh}, \citenamefont {Devi},\ and\ \citenamefont {Rajagopal}}]{MI}%
  \BibitemOpen
  \bibfield  {author} {\bibinfo {author} {\bibfnamefont {H.~S.}\ \bibnamefont
  {Karthik}}, \bibinfo {author} {\bibfnamefont {H.}~\bibnamefont {Katiyar}},
  \bibinfo {author} {\bibfnamefont {A.}~\bibnamefont {Shukla}}, \bibinfo
  {author} {\bibfnamefont {T.~S.}\ \bibnamefont {Mahesh}}, \bibinfo {author}
  {\bibfnamefont {A.~R.~U.}\ \bibnamefont {Devi}}, \ and\ \bibinfo {author}
  {\bibfnamefont {A.~K.}\ \bibnamefont {Rajagopal}},\ }\href {\doibase
  10.1103/PhysRevA.87.052118} {\bibfield  {journal} {\bibinfo  {journal} {Phys.
  Rev. A}\ }\textbf {\bibinfo {volume} {87}},\ \bibinfo {pages} {052118}
  (\bibinfo {year} {2013})}\BibitemShut {NoStop}%
\bibitem [{\citenamefont {Moussa}\ \emph {et~al.}(2010)\citenamefont {Moussa},
  \citenamefont {Ryan}, \citenamefont {Cory},\ and\ \citenamefont
  {Laflamme}}]{moussa1qbitcontextuality2010}%
  \BibitemOpen
  \bibfield  {author} {\bibinfo {author} {\bibfnamefont {O.}~\bibnamefont
  {Moussa}}, \bibinfo {author} {\bibfnamefont {C.~A.}\ \bibnamefont {Ryan}},
  \bibinfo {author} {\bibfnamefont {D.~G.}\ \bibnamefont {Cory}}, \ and\
  \bibinfo {author} {\bibfnamefont {R.}~\bibnamefont {Laflamme}},\ }\href
  {\doibase 10.1103/PhysRevLett.104.160501} {\bibfield  {journal} {\bibinfo
  {journal} {Phys. Rev. Lett.}\ }\textbf {\bibinfo {volume} {104}},\ \bibinfo
  {pages} {160501} (\bibinfo {year} {2010})}\BibitemShut {NoStop}%
\bibitem [{\citenamefont {Joshi}\ \emph {et~al.}(2014)\citenamefont {Joshi},
  \citenamefont {Shukla}, \citenamefont {Katiyar}, \citenamefont {Hazra},\ and\
  \citenamefont {Mahesh}}]{joshi2014estimating}%
  \BibitemOpen
  \bibfield  {author} {\bibinfo {author} {\bibfnamefont {S.}~\bibnamefont
  {Joshi}}, \bibinfo {author} {\bibfnamefont {A.}~\bibnamefont {Shukla}},
  \bibinfo {author} {\bibfnamefont {H.}~\bibnamefont {Katiyar}}, \bibinfo
  {author} {\bibfnamefont {A.}~\bibnamefont {Hazra}}, \ and\ \bibinfo {author}
  {\bibfnamefont {T.~S.}\ \bibnamefont {Mahesh}},\ }\href@noop {} {\bibfield
  {journal} {\bibinfo  {journal} {Physical Review A}\ }\textbf {\bibinfo
  {volume} {90}},\ \bibinfo {pages} {022303} (\bibinfo {year}
  {2014})}\BibitemShut {NoStop}%
\bibitem [{\citenamefont {Demtroder}(2006)}]{demtroder2006atoms}%
  \BibitemOpen
  \bibfield  {author} {\bibinfo {author} {\bibfnamefont {W.}~\bibnamefont
  {Demtroder}},\ }\href@noop {} {\bibfield  {journal} {\bibinfo  {journal}
  {Heidelberg, Springer-Verlag Berlin Heidelberg, 2006. 571}\ } (\bibinfo
  {year} {2006})}\BibitemShut {NoStop}%
\bibitem [{\citenamefont {Bowers}\ \emph {et~al.}(1984)\citenamefont {Bowers},
  \citenamefont {Delbert}, \citenamefont {Hunter},\ and\ \citenamefont
  {McIver~Jr}}]{bowers1984fragmentation}%
  \BibitemOpen
  \bibfield  {author} {\bibinfo {author} {\bibfnamefont {W.~D.}\ \bibnamefont
  {Bowers}}, \bibinfo {author} {\bibfnamefont {S.~S.}\ \bibnamefont {Delbert}},
  \bibinfo {author} {\bibfnamefont {R.~L.}\ \bibnamefont {Hunter}}, \ and\
  \bibinfo {author} {\bibfnamefont {R.~T.}\ \bibnamefont {McIver~Jr}},\
  }\href@noop {} {\bibfield  {journal} {\bibinfo  {journal} {Journal of the
  American Chemical Society}\ }\textbf {\bibinfo {volume} {106}},\ \bibinfo
  {pages} {7288} (\bibinfo {year} {1984})}\BibitemShut {NoStop}%
\bibitem [{\citenamefont {Peres}(1990)}]{peres_context_1pg}%
  \BibitemOpen
  \bibfield  {author} {\bibinfo {author} {\bibfnamefont {A.}~\bibnamefont
  {Peres}},\ }\href
  {http://www.sciencedirect.com/science/article/pii/037596019090172K}
  {\bibfield  {journal} {\bibinfo  {journal} {Physics Letters A}\ }\textbf
  {\bibinfo {volume} {151}},\ \bibinfo {pages} {107 } (\bibinfo {year}
  {1990})}\BibitemShut {NoStop}%
\bibitem [{\citenamefont {Nielsen}\ and\ \citenamefont
  {Chuang}(2010{\natexlab{a}})}]{quant_info_neilson_chuang}%
  \BibitemOpen
  \bibfield  {author} {\bibinfo {author} {\bibfnamefont {M.~A.}\ \bibnamefont
  {Nielsen}}\ and\ \bibinfo {author} {\bibfnamefont {I.~L.}\ \bibnamefont
  {Chuang}},\ }\href {http://dx.doi.org/10.1017/CBO9780511976667} {\emph
  {\bibinfo {title} {Quantum Computation and Quantum Information}}}\ (\bibinfo
  {publisher} {Cambridge University Press},\ \bibinfo {year} {2010})\ \bibinfo
  {note} {cambridge Books Online}\BibitemShut {NoStop}%
\bibitem [{\citenamefont {Su}\ \emph {et~al.}(2012)\citenamefont {Su},
  \citenamefont {Chen}, \citenamefont {Wu}, \citenamefont {Yu},\ and\
  \citenamefont {Oh}}]{Cont_theory}%
  \BibitemOpen
  \bibfield  {author} {\bibinfo {author} {\bibfnamefont {H.-Y.}\ \bibnamefont
  {Su}}, \bibinfo {author} {\bibfnamefont {J.-L.}\ \bibnamefont {Chen}},
  \bibinfo {author} {\bibfnamefont {C.}~\bibnamefont {Wu}}, \bibinfo {author}
  {\bibfnamefont {S.}~\bibnamefont {Yu}}, \ and\ \bibinfo {author}
  {\bibfnamefont {C.~H.}\ \bibnamefont {Oh}},\ }\href {\doibase
  10.1103/PhysRevA.85.052126} {\bibfield  {journal} {\bibinfo  {journal} {Phys.
  Rev. A}\ }\textbf {\bibinfo {volume} {85}},\ \bibinfo {pages} {052126}
  (\bibinfo {year} {2012})}\BibitemShut {NoStop}%
\bibitem [{\citenamefont {Katiyar}\ \emph {et~al.}(2015)\citenamefont
  {Katiyar}, \citenamefont {Kumar},\ and\ \citenamefont
  {Mahesh}}]{katiyar2015investigation}%
  \BibitemOpen
  \bibfield  {author} {\bibinfo {author} {\bibfnamefont {H.}~\bibnamefont
  {Katiyar}}, \bibinfo {author} {\bibfnamefont {C.}~\bibnamefont {Kumar}}, \
  and\ \bibinfo {author} {\bibfnamefont {T.~S.}\ \bibnamefont {Mahesh}},\
  }\href@noop {} {\bibfield  {journal} {\bibinfo  {journal} {arXiv preprint
  arXiv:1503.05883}\ } (\bibinfo {year} {2015})}\BibitemShut {NoStop}%
\bibitem [{\citenamefont {Nielsen}\ and\ \citenamefont
  {Chuang}(2010{\natexlab{b}})}]{chuangbook}%
  \BibitemOpen
  \bibfield  {author} {\bibinfo {author} {\bibnamefont {Nielsen}}\ and\
  \bibinfo {author} {\bibfnamefont {I.~L.}\ \bibnamefont {Chuang}},\
  }\href@noop {} {\emph {\bibinfo {title} {Quantum computation and quantum
  information}}}\ (\bibinfo  {publisher} {Cambridge university press},\
  \bibinfo {year} {2010})\BibitemShut {NoStop}%
\bibitem [{\citenamefont {I.~L.~Chuang}\ and\ \citenamefont
  {Leung}(1998)}]{ChuangPRSL98}%
  \BibitemOpen
  \bibfield  {author} {\bibinfo {author} {\bibfnamefont {M.~G.~K.}\
  \bibnamefont {I.~L.~Chuang}, \bibfnamefont {N.~Gershenfeld}}\ and\ \bibinfo
  {author} {\bibfnamefont {D.~W.}\ \bibnamefont {Leung}},\ }\href@noop {}
  {\bibfield  {journal} {\bibinfo  {journal} {Proc. R. Soc. Lond.,Ser A}\
  }\textbf {\bibinfo {volume} {454}},\ \bibinfo {pages} {447} (\bibinfo {year}
  {1998})}\BibitemShut {NoStop}%
\bibitem [{\citenamefont {Leung}\ \emph {et~al.}(1999)\citenamefont {Leung},
  \citenamefont {Vandersypen}, \citenamefont {Zhou}, \citenamefont {Sherwood},
  \citenamefont {Yannoni}, \citenamefont {Kubinec},\ and\ \citenamefont
  {Chuang}}]{ChuangPRA99}%
  \BibitemOpen
  \bibfield  {author} {\bibinfo {author} {\bibfnamefont {D.}~\bibnamefont
  {Leung}}, \bibinfo {author} {\bibfnamefont {L.}~\bibnamefont {Vandersypen}},
  \bibinfo {author} {\bibfnamefont {X.}~\bibnamefont {Zhou}}, \bibinfo {author}
  {\bibfnamefont {M.}~\bibnamefont {Sherwood}}, \bibinfo {author}
  {\bibfnamefont {C.}~\bibnamefont {Yannoni}}, \bibinfo {author} {\bibfnamefont
  {M.}~\bibnamefont {Kubinec}}, \ and\ \bibinfo {author} {\bibfnamefont
  {I.}~\bibnamefont {Chuang}},\ }\href@noop {} {\bibfield  {journal} {\bibinfo
  {journal} {Physical Review A}\ }\textbf {\bibinfo {volume} {60}},\ \bibinfo
  {pages} {1924} (\bibinfo {year} {1999})}\BibitemShut {NoStop}%
\bibitem [{\citenamefont {Das}\ \emph {et~al.}(2003)\citenamefont {Das},
  \citenamefont {Mahesh},\ and\ \citenamefont {Kumar}}]{Aniltomo}%
  \BibitemOpen
  \bibfield  {author} {\bibinfo {author} {\bibfnamefont {R.}~\bibnamefont
  {Das}}, \bibinfo {author} {\bibfnamefont {T.~S.}\ \bibnamefont {Mahesh}}, \
  and\ \bibinfo {author} {\bibfnamefont {A.}~\bibnamefont {Kumar}},\ }\href
  {\doibase 10.1103/PhysRevA.67.062304} {\bibfield  {journal} {\bibinfo
  {journal} {Phys. Rev. A}\ }\textbf {\bibinfo {volume} {67}},\ \bibinfo
  {pages} {062304} (\bibinfo {year} {2003})}\BibitemShut {NoStop}%
\bibitem [{\citenamefont {Allahverdyan}\ \emph {et~al.}(2004)\citenamefont
  {Allahverdyan}, \citenamefont {Balian},\ and\ \citenamefont
  {Nieuwenhuizen}}]{Nieuwenhuizen}%
  \BibitemOpen
  \bibfield  {author} {\bibinfo {author} {\bibfnamefont {A.~E.}\ \bibnamefont
  {Allahverdyan}}, \bibinfo {author} {\bibfnamefont {R.}~\bibnamefont
  {Balian}}, \ and\ \bibinfo {author} {\bibfnamefont {T.~M.}\ \bibnamefont
  {Nieuwenhuizen}},\ }\href {\doibase 10.1103/PhysRevLett.92.120402} {\bibfield
   {journal} {\bibinfo  {journal} {Phys. Rev. Lett.}\ }\textbf {\bibinfo
  {volume} {92}},\ \bibinfo {pages} {120402} (\bibinfo {year}
  {2004})}\BibitemShut {NoStop}%
\bibitem [{\citenamefont {Peng}\ \emph {et~al.}(2007)\citenamefont {Peng},
  \citenamefont {Du},\ and\ \citenamefont {Suter}}]{sutertomo}%
  \BibitemOpen
  \bibfield  {author} {\bibinfo {author} {\bibfnamefont {X.}~\bibnamefont
  {Peng}}, \bibinfo {author} {\bibfnamefont {J.}~\bibnamefont {Du}}, \ and\
  \bibinfo {author} {\bibfnamefont {D.}~\bibnamefont {Suter}},\ }\href
  {\doibase 10.1103/PhysRevA.76.042117} {\bibfield  {journal} {\bibinfo
  {journal} {Phys. Rev. A}\ }\textbf {\bibinfo {volume} {76}},\ \bibinfo
  {pages} {042117} (\bibinfo {year} {2007})}\BibitemShut {NoStop}%
\bibitem [{\citenamefont {Shukla}\ \emph {et~al.}(2013)\citenamefont {Shukla},
  \citenamefont {Rao},\ and\ \citenamefont {Mahesh}}]{abhishekQST2013}%
  \BibitemOpen
  \bibfield  {author} {\bibinfo {author} {\bibfnamefont {A.}~\bibnamefont
  {Shukla}}, \bibinfo {author} {\bibfnamefont {K.~R.~K.}\ \bibnamefont {Rao}},
  \ and\ \bibinfo {author} {\bibfnamefont {T.~S.}\ \bibnamefont {Mahesh}},\
  }\href {\doibase 10.1103/PhysRevA.87.062317} {\bibfield  {journal} {\bibinfo
  {journal} {Phys. Rev. A}\ }\textbf {\bibinfo {volume} {87}},\ \bibinfo
  {pages} {062317} (\bibinfo {year} {2013})}\BibitemShut {NoStop}%
\bibitem [{\citenamefont {Chuang}\ and\ \citenamefont
  {Nielsen}(1997)}]{chuang97}%
  \BibitemOpen
  \bibfield  {author} {\bibinfo {author} {\bibfnamefont {I.~L.}\ \bibnamefont
  {Chuang}}\ and\ \bibinfo {author} {\bibfnamefont {M.~A.}\ \bibnamefont
  {Nielsen}},\ }\href@noop {} {\bibfield  {journal} {\bibinfo  {journal}
  {Journal of Modern Optics}\ }\textbf {\bibinfo {volume} {44}},\ \bibinfo
  {pages} {2455} (\bibinfo {year} {1997})}\BibitemShut {NoStop}%
\bibitem [{\citenamefont {Poyatos}\ \emph {et~al.}(1997)\citenamefont
  {Poyatos}, \citenamefont {Cirac},\ and\ \citenamefont {Zoller}}]{zollar97}%
  \BibitemOpen
  \bibfield  {author} {\bibinfo {author} {\bibfnamefont {J.~F.}\ \bibnamefont
  {Poyatos}}, \bibinfo {author} {\bibfnamefont {J.~I.}\ \bibnamefont {Cirac}},
  \ and\ \bibinfo {author} {\bibfnamefont {P.}~\bibnamefont {Zoller}},\ }\href
  {\doibase 10.1103/PhysRevLett.78.390} {\bibfield  {journal} {\bibinfo
  {journal} {Phys. Rev. Lett.}\ }\textbf {\bibinfo {volume} {78}},\ \bibinfo
  {pages} {390} (\bibinfo {year} {1997})}\BibitemShut {NoStop}%
\bibitem [{\citenamefont {Childs}\ \emph {et~al.}(2001)\citenamefont {Childs},
  \citenamefont {Chuang},\ and\ \citenamefont {Leung}}]{ChuangPRA2001}%
  \BibitemOpen
  \bibfield  {author} {\bibinfo {author} {\bibfnamefont {A.~M.}\ \bibnamefont
  {Childs}}, \bibinfo {author} {\bibfnamefont {I.~L.}\ \bibnamefont {Chuang}},
  \ and\ \bibinfo {author} {\bibfnamefont {D.~W.}\ \bibnamefont {Leung}},\
  }\href {\doibase 10.1103/PhysRevA.64.012314} {\bibfield  {journal} {\bibinfo
  {journal} {Phys. Rev. A}\ }\textbf {\bibinfo {volume} {64}},\ \bibinfo
  {pages} {012314} (\bibinfo {year} {2001})}\BibitemShut {NoStop}%
\bibitem [{\citenamefont {Weinstein}\ \emph {et~al.}(2004)\citenamefont
  {Weinstein}, \citenamefont {Havel}, \citenamefont {Emerson}, \citenamefont
  {Boulant}, \citenamefont {Saraceno}, \citenamefont {Lloyd},\ and\
  \citenamefont {Cory}}]{QPTofQFT}%
  \BibitemOpen
  \bibfield  {author} {\bibinfo {author} {\bibfnamefont {Y.~S.}\ \bibnamefont
  {Weinstein}}, \bibinfo {author} {\bibfnamefont {T.~F.}\ \bibnamefont
  {Havel}}, \bibinfo {author} {\bibfnamefont {J.}~\bibnamefont {Emerson}},
  \bibinfo {author} {\bibfnamefont {N.}~\bibnamefont {Boulant}}, \bibinfo
  {author} {\bibfnamefont {M.}~\bibnamefont {Saraceno}}, \bibinfo {author}
  {\bibfnamefont {S.}~\bibnamefont {Lloyd}}, \ and\ \bibinfo {author}
  {\bibfnamefont {D.~G.}\ \bibnamefont {Cory}},\ }\href {\doibase
  http://dx.doi.org/10.1063/1.1785151} {\bibfield  {journal} {\bibinfo
  {journal} {The Journal of Chemical Physics}\ }\textbf {\bibinfo {volume}
  {121}},\ \bibinfo {pages} {6117} (\bibinfo {year} {2004})}\BibitemShut
  {NoStop}%
\bibitem [{\citenamefont {De~Martini}\ \emph {et~al.}(2003)\citenamefont
  {De~Martini}, \citenamefont {Mazzei}, \citenamefont {Ricci},\ and\
  \citenamefont {D'Ariano}}]{EAPT1Exp}%
  \BibitemOpen
  \bibfield  {author} {\bibinfo {author} {\bibfnamefont {F.}~\bibnamefont
  {De~Martini}}, \bibinfo {author} {\bibfnamefont {A.}~\bibnamefont {Mazzei}},
  \bibinfo {author} {\bibfnamefont {M.}~\bibnamefont {Ricci}}, \ and\ \bibinfo
  {author} {\bibfnamefont {G.~M.}\ \bibnamefont {D'Ariano}},\ }\href {\doibase
  10.1103/PhysRevA.67.062307} {\bibfield  {journal} {\bibinfo  {journal} {Phys.
  Rev. A}\ }\textbf {\bibinfo {volume} {67}},\ \bibinfo {pages} {062307}
  (\bibinfo {year} {2003})}\BibitemShut {NoStop}%
\bibitem [{\citenamefont {Altepeter}\ \emph {et~al.}(2003)\citenamefont
  {Altepeter}, \citenamefont {Branning}, \citenamefont {Jeffrey}, \citenamefont
  {Wei}, \citenamefont {Kwiat}, \citenamefont {Thew}, \citenamefont {O'Brien},
  \citenamefont {Nielsen},\ and\ \citenamefont {White}}]{altepeter}%
  \BibitemOpen
  \bibfield  {author} {\bibinfo {author} {\bibfnamefont {J.~B.}\ \bibnamefont
  {Altepeter}}, \bibinfo {author} {\bibfnamefont {D.}~\bibnamefont {Branning}},
  \bibinfo {author} {\bibfnamefont {E.}~\bibnamefont {Jeffrey}}, \bibinfo
  {author} {\bibfnamefont {T.~C.}\ \bibnamefont {Wei}}, \bibinfo {author}
  {\bibfnamefont {P.~G.}\ \bibnamefont {Kwiat}}, \bibinfo {author}
  {\bibfnamefont {R.~T.}\ \bibnamefont {Thew}}, \bibinfo {author}
  {\bibfnamefont {J.~L.}\ \bibnamefont {O'Brien}}, \bibinfo {author}
  {\bibfnamefont {M.~A.}\ \bibnamefont {Nielsen}}, \ and\ \bibinfo {author}
  {\bibfnamefont {A.~G.}\ \bibnamefont {White}},\ }\href {\doibase
  10.1103/PhysRevLett.90.193601} {\bibfield  {journal} {\bibinfo  {journal}
  {Phys. Rev. Lett.}\ }\textbf {\bibinfo {volume} {90}},\ \bibinfo {pages}
  {193601} (\bibinfo {year} {2003})}\BibitemShut {NoStop}%
\bibitem [{\citenamefont {O'Brien}\ \emph {et~al.}(2004)\citenamefont
  {O'Brien}, \citenamefont {Pryde}, \citenamefont {Gilchrist}, \citenamefont
  {James}, \citenamefont {Langford}, \citenamefont {Ralph},\ and\ \citenamefont
  {White}}]{obrian}%
  \BibitemOpen
  \bibfield  {author} {\bibinfo {author} {\bibfnamefont {J.~L.}\ \bibnamefont
  {O'Brien}}, \bibinfo {author} {\bibfnamefont {G.~J.}\ \bibnamefont {Pryde}},
  \bibinfo {author} {\bibfnamefont {A.}~\bibnamefont {Gilchrist}}, \bibinfo
  {author} {\bibfnamefont {D.~F.~V.}\ \bibnamefont {James}}, \bibinfo {author}
  {\bibfnamefont {N.~K.}\ \bibnamefont {Langford}}, \bibinfo {author}
  {\bibfnamefont {T.~C.}\ \bibnamefont {Ralph}}, \ and\ \bibinfo {author}
  {\bibfnamefont {A.~G.}\ \bibnamefont {White}},\ }\href {\doibase
  10.1103/PhysRevLett.93.080502} {\bibfield  {journal} {\bibinfo  {journal}
  {Phys. Rev. Lett.}\ }\textbf {\bibinfo {volume} {93}},\ \bibinfo {pages}
  {080502} (\bibinfo {year} {2004})}\BibitemShut {NoStop}%
\bibitem [{\citenamefont {Mitchell}\ \emph {et~al.}(2003)\citenamefont
  {Mitchell}, \citenamefont {Ellenor}, \citenamefont {Schneider},\ and\
  \citenamefont {Steinberg}}]{bellstatefilter}%
  \BibitemOpen
  \bibfield  {author} {\bibinfo {author} {\bibfnamefont {M.~W.}\ \bibnamefont
  {Mitchell}}, \bibinfo {author} {\bibfnamefont {C.~W.}\ \bibnamefont
  {Ellenor}}, \bibinfo {author} {\bibfnamefont {S.}~\bibnamefont {Schneider}},
  \ and\ \bibinfo {author} {\bibfnamefont {A.~M.}\ \bibnamefont {Steinberg}},\
  }\href {\doibase 10.1103/PhysRevLett.91.120402} {\bibfield  {journal}
  {\bibinfo  {journal} {Phys. Rev. Lett.}\ }\textbf {\bibinfo {volume} {91}},\
  \bibinfo {pages} {120402} (\bibinfo {year} {2003})}\BibitemShut {NoStop}%
\bibitem [{\citenamefont {Riebe}\ \emph {et~al.}(2006)\citenamefont {Riebe},
  \citenamefont {Kim}, \citenamefont {Schindler}, \citenamefont {Monz},
  \citenamefont {Schmidt}, \citenamefont {K\"orber}, \citenamefont {H\"ansel},
  \citenamefont {H\"affner}, \citenamefont {Roos},\ and\ \citenamefont
  {Blatt}}]{qptITprl2006}%
  \BibitemOpen
  \bibfield  {author} {\bibinfo {author} {\bibfnamefont {M.}~\bibnamefont
  {Riebe}}, \bibinfo {author} {\bibfnamefont {K.}~\bibnamefont {Kim}}, \bibinfo
  {author} {\bibfnamefont {P.}~\bibnamefont {Schindler}}, \bibinfo {author}
  {\bibfnamefont {T.}~\bibnamefont {Monz}}, \bibinfo {author} {\bibfnamefont
  {P.~O.}\ \bibnamefont {Schmidt}}, \bibinfo {author} {\bibfnamefont {T.~K.}\
  \bibnamefont {K\"orber}}, \bibinfo {author} {\bibfnamefont {W.}~\bibnamefont
  {H\"ansel}}, \bibinfo {author} {\bibfnamefont {H.}~\bibnamefont {H\"affner}},
  \bibinfo {author} {\bibfnamefont {C.~F.}\ \bibnamefont {Roos}}, \ and\
  \bibinfo {author} {\bibfnamefont {R.}~\bibnamefont {Blatt}},\ }\href
  {\doibase 10.1103/PhysRevLett.97.220407} {\bibfield  {journal} {\bibinfo
  {journal} {Phys. Rev. Lett.}\ }\textbf {\bibinfo {volume} {97}},\ \bibinfo
  {pages} {220407} (\bibinfo {year} {2006})}\BibitemShut {NoStop}%
\bibitem [{\citenamefont {Hanneke}\ \emph {et~al.}(2010)\citenamefont
  {Hanneke}, \citenamefont {Home}, \citenamefont {Jost}, \citenamefont {Amini},
  \citenamefont {Leibfried},\ and\ \citenamefont
  {Wineland}}]{QPT_IonTrapNature2010}%
  \BibitemOpen
  \bibfield  {author} {\bibinfo {author} {\bibfnamefont {D.}~\bibnamefont
  {Hanneke}}, \bibinfo {author} {\bibfnamefont {J.~P.}\ \bibnamefont {Home}},
  \bibinfo {author} {\bibfnamefont {J.~D.}\ \bibnamefont {Jost}}, \bibinfo
  {author} {\bibfnamefont {J.~M.}\ \bibnamefont {Amini}}, \bibinfo {author}
  {\bibfnamefont {D.}~\bibnamefont {Leibfried}}, \ and\ \bibinfo {author}
  {\bibfnamefont {D.~J.}\ \bibnamefont {Wineland}},\ }\href {\doibase
  10.1038/nphys1453} {\bibfield  {journal} {\bibinfo  {journal} {Nature Phys.}\
  }\textbf {\bibinfo {volume} {6}},\ \bibinfo {pages} {13} (\bibinfo {year}
  {2010})}\BibitemShut {NoStop}%
\bibitem [{\citenamefont {Neeley}\ \emph {et~al.}(2008)\citenamefont {Neeley},
  \citenamefont {Ansmann}, \citenamefont {Bialczak}, \citenamefont {Hofheinz},
  \citenamefont {Katz}, \citenamefont {Lucero}, \citenamefont {O/'Connell},
  \citenamefont {Wang}, \citenamefont {Cleland},\ and\ \citenamefont
  {Martinis}}]{QPT_SQUID}%
  \BibitemOpen
  \bibfield  {author} {\bibinfo {author} {\bibfnamefont {M.}~\bibnamefont
  {Neeley}}, \bibinfo {author} {\bibfnamefont {M.}~\bibnamefont {Ansmann}},
  \bibinfo {author} {\bibfnamefont {R.~C.}\ \bibnamefont {Bialczak}}, \bibinfo
  {author} {\bibfnamefont {M.}~\bibnamefont {Hofheinz}}, \bibinfo {author}
  {\bibfnamefont {N.}~\bibnamefont {Katz}}, \bibinfo {author} {\bibfnamefont
  {E.}~\bibnamefont {Lucero}}, \bibinfo {author} {\bibfnamefont
  {A.}~\bibnamefont {O/'Connell}}, \bibinfo {author} {\bibfnamefont
  {H.}~\bibnamefont {Wang}}, \bibinfo {author} {\bibfnamefont {A.~N.}\
  \bibnamefont {Cleland}}, \ and\ \bibinfo {author} {\bibfnamefont {J.~M.}\
  \bibnamefont {Martinis}},\ }\href {\doibase 10.1038/nphys972} {\bibfield
  {journal} {\bibinfo  {journal} {Nature Phys.}\ }\textbf {\bibinfo {volume}
  {4}},\ \bibinfo {pages} {523} (\bibinfo {year} {2008})}\BibitemShut {NoStop}%
\bibitem [{\citenamefont {Chow}\ \emph {et~al.}(2009)\citenamefont {Chow},
  \citenamefont {Gambetta}, \citenamefont {Tornberg}, \citenamefont {Koch},
  \citenamefont {Bishop}, \citenamefont {Houck}, \citenamefont {Johnson},
  \citenamefont {Frunzio}, \citenamefont {Girvin},\ and\ \citenamefont
  {Schoelkopf}}]{SQUID2009}%
  \BibitemOpen
  \bibfield  {author} {\bibinfo {author} {\bibfnamefont {J.~M.}\ \bibnamefont
  {Chow}}, \bibinfo {author} {\bibfnamefont {J.~M.}\ \bibnamefont {Gambetta}},
  \bibinfo {author} {\bibfnamefont {L.}~\bibnamefont {Tornberg}}, \bibinfo
  {author} {\bibfnamefont {J.}~\bibnamefont {Koch}}, \bibinfo {author}
  {\bibfnamefont {L.~S.}\ \bibnamefont {Bishop}}, \bibinfo {author}
  {\bibfnamefont {A.~A.}\ \bibnamefont {Houck}}, \bibinfo {author}
  {\bibfnamefont {B.~R.}\ \bibnamefont {Johnson}}, \bibinfo {author}
  {\bibfnamefont {L.}~\bibnamefont {Frunzio}}, \bibinfo {author} {\bibfnamefont
  {S.~M.}\ \bibnamefont {Girvin}}, \ and\ \bibinfo {author} {\bibfnamefont
  {R.~J.}\ \bibnamefont {Schoelkopf}},\ }\href {\doibase
  10.1103/PhysRevLett.102.090502} {\bibfield  {journal} {\bibinfo  {journal}
  {Phys. Rev. Lett.}\ }\textbf {\bibinfo {volume} {102}},\ \bibinfo {pages}
  {090502} (\bibinfo {year} {2009})}\BibitemShut {NoStop}%
\bibitem [{\citenamefont {Bialczak}\ \emph {et~al.}(2010)\citenamefont
  {Bialczak}, \citenamefont {Ansmann}, \citenamefont {Hofheinz}, \citenamefont
  {Lucero}, \citenamefont {Neeley}, \citenamefont {O/'Connell}, \citenamefont
  {Sank}, \citenamefont {Wang}, \citenamefont {Wenner}, \citenamefont
  {Steffen},\ and\ \citenamefont {Cleland}}]{MartiniSQUID2010}%
  \BibitemOpen
  \bibfield  {author} {\bibinfo {author} {\bibfnamefont {R.~C.}\ \bibnamefont
  {Bialczak}}, \bibinfo {author} {\bibfnamefont {M.}~\bibnamefont {Ansmann}},
  \bibinfo {author} {\bibfnamefont {M.}~\bibnamefont {Hofheinz}}, \bibinfo
  {author} {\bibfnamefont {E.}~\bibnamefont {Lucero}}, \bibinfo {author}
  {\bibfnamefont {M.}~\bibnamefont {Neeley}}, \bibinfo {author} {\bibfnamefont
  {A.~D.}\ \bibnamefont {O/'Connell}}, \bibinfo {author} {\bibfnamefont
  {D.}~\bibnamefont {Sank}}, \bibinfo {author} {\bibfnamefont {H.}~\bibnamefont
  {Wang}}, \bibinfo {author} {\bibfnamefont {J.}~\bibnamefont {Wenner}},
  \bibinfo {author} {\bibfnamefont {M.}~\bibnamefont {Steffen}}, \ and\
  \bibinfo {author} {\bibfnamefont {J.~M.}\ \bibnamefont {Cleland},
  \bibfnamefont {A.~N.and~Martinis}},\ }\href {\doibase 10.1038/nphys1639}
  {\bibfield  {journal} {\bibinfo  {journal} {Nature Phys.}\ }\textbf {\bibinfo
  {volume} {6}},\ \bibinfo {pages} {409} (\bibinfo {year} {2010})}\BibitemShut
  {NoStop}%
\bibitem [{\citenamefont {Yamamoto}\ \emph {et~al.}(2010)\citenamefont
  {Yamamoto}, \citenamefont {Neeley}, \citenamefont {Lucero}, \citenamefont
  {Bialczak}, \citenamefont {Kelly}, \citenamefont {Lenander}, \citenamefont
  {Mariantoni}, \citenamefont {O'Connell}, \citenamefont {Sank}, \citenamefont
  {Wang}, \citenamefont {Weides}, \citenamefont {Wenner}, \citenamefont {Yin},
  \citenamefont {Cleland},\ and\ \citenamefont {Martinis}}]{QPT2spSQUID}%
  \BibitemOpen
  \bibfield  {author} {\bibinfo {author} {\bibfnamefont {T.}~\bibnamefont
  {Yamamoto}}, \bibinfo {author} {\bibfnamefont {M.}~\bibnamefont {Neeley}},
  \bibinfo {author} {\bibfnamefont {E.}~\bibnamefont {Lucero}}, \bibinfo
  {author} {\bibfnamefont {R.~C.}\ \bibnamefont {Bialczak}}, \bibinfo {author}
  {\bibfnamefont {J.}~\bibnamefont {Kelly}}, \bibinfo {author} {\bibfnamefont
  {M.}~\bibnamefont {Lenander}}, \bibinfo {author} {\bibfnamefont
  {M.}~\bibnamefont {Mariantoni}}, \bibinfo {author} {\bibfnamefont {A.~D.}\
  \bibnamefont {O'Connell}}, \bibinfo {author} {\bibfnamefont {D.}~\bibnamefont
  {Sank}}, \bibinfo {author} {\bibfnamefont {H.}~\bibnamefont {Wang}}, \bibinfo
  {author} {\bibfnamefont {M.}~\bibnamefont {Weides}}, \bibinfo {author}
  {\bibfnamefont {J.}~\bibnamefont {Wenner}}, \bibinfo {author} {\bibfnamefont
  {Y.}~\bibnamefont {Yin}}, \bibinfo {author} {\bibfnamefont {A.~N.}\
  \bibnamefont {Cleland}}, \ and\ \bibinfo {author} {\bibfnamefont {J.~M.}\
  \bibnamefont {Martinis}},\ }\href {\doibase 10.1103/PhysRevB.82.184515}
  {\bibfield  {journal} {\bibinfo  {journal} {Phys. Rev. B}\ }\textbf {\bibinfo
  {volume} {82}},\ \bibinfo {pages} {184515} (\bibinfo {year}
  {2010})}\BibitemShut {NoStop}%
\bibitem [{\citenamefont {Chow}\ \emph {et~al.}(2011)\citenamefont {Chow},
  \citenamefont {C\'orcoles}, \citenamefont {Gambetta}, \citenamefont
  {Rigetti}, \citenamefont {Johnson}, \citenamefont {Smolin}, \citenamefont
  {Rozen}, \citenamefont {Keefe}, \citenamefont {Rothwell}, \citenamefont
  {Ketchen},\ and\ \citenamefont {Steffen}}]{QPT_SQUIDChow2011}%
  \BibitemOpen
  \bibfield  {author} {\bibinfo {author} {\bibfnamefont {J.~M.}\ \bibnamefont
  {Chow}}, \bibinfo {author} {\bibfnamefont {A.~D.}\ \bibnamefont
  {C\'orcoles}}, \bibinfo {author} {\bibfnamefont {J.~M.}\ \bibnamefont
  {Gambetta}}, \bibinfo {author} {\bibfnamefont {C.}~\bibnamefont {Rigetti}},
  \bibinfo {author} {\bibfnamefont {B.~R.}\ \bibnamefont {Johnson}}, \bibinfo
  {author} {\bibfnamefont {J.~A.}\ \bibnamefont {Smolin}}, \bibinfo {author}
  {\bibfnamefont {J.~R.}\ \bibnamefont {Rozen}}, \bibinfo {author}
  {\bibfnamefont {G.~A.}\ \bibnamefont {Keefe}}, \bibinfo {author}
  {\bibfnamefont {M.~B.}\ \bibnamefont {Rothwell}}, \bibinfo {author}
  {\bibfnamefont {M.~B.}\ \bibnamefont {Ketchen}}, \ and\ \bibinfo {author}
  {\bibfnamefont {M.}~\bibnamefont {Steffen}},\ }\href {\doibase
  10.1103/PhysRevLett.107.080502} {\bibfield  {journal} {\bibinfo  {journal}
  {Phys. Rev. Lett.}\ }\textbf {\bibinfo {volume} {107}},\ \bibinfo {pages}
  {080502} (\bibinfo {year} {2011})}\BibitemShut {NoStop}%
\bibitem [{\citenamefont {Dewes}\ \emph {et~al.}(2012)\citenamefont {Dewes},
  \citenamefont {Ong}, \citenamefont {Schmitt}, \citenamefont {Lauro},
  \citenamefont {Boulant}, \citenamefont {Bertet}, \citenamefont {Vion},\ and\
  \citenamefont {Esteve}}]{SQUID_Dewes2012}%
  \BibitemOpen
  \bibfield  {author} {\bibinfo {author} {\bibfnamefont {A.}~\bibnamefont
  {Dewes}}, \bibinfo {author} {\bibfnamefont {F.~R.}\ \bibnamefont {Ong}},
  \bibinfo {author} {\bibfnamefont {V.}~\bibnamefont {Schmitt}}, \bibinfo
  {author} {\bibfnamefont {R.}~\bibnamefont {Lauro}}, \bibinfo {author}
  {\bibfnamefont {N.}~\bibnamefont {Boulant}}, \bibinfo {author} {\bibfnamefont
  {P.}~\bibnamefont {Bertet}}, \bibinfo {author} {\bibfnamefont
  {D.}~\bibnamefont {Vion}}, \ and\ \bibinfo {author} {\bibfnamefont
  {D.}~\bibnamefont {Esteve}},\ }\href {\doibase
  10.1103/PhysRevLett.108.057002} {\bibfield  {journal} {\bibinfo  {journal}
  {Phys. Rev. Lett.}\ }\textbf {\bibinfo {volume} {108}},\ \bibinfo {pages}
  {057002} (\bibinfo {year} {2012})}\BibitemShut {NoStop}%
\bibitem [{\citenamefont {Zhang}\ \emph {et~al.}(2014)\citenamefont {Zhang},
  \citenamefont {Souza}, \citenamefont {Brandao},\ and\ \citenamefont
  {Suter}}]{suterprotectedgate}%
  \BibitemOpen
  \bibfield  {author} {\bibinfo {author} {\bibfnamefont {J.}~\bibnamefont
  {Zhang}}, \bibinfo {author} {\bibfnamefont {A.~M.}\ \bibnamefont {Souza}},
  \bibinfo {author} {\bibfnamefont {F.~D.}\ \bibnamefont {Brandao}}, \ and\
  \bibinfo {author} {\bibfnamefont {D.}~\bibnamefont {Suter}},\ }\href
  {\doibase 10.1103/PhysRevLett.112.050502} {\bibfield  {journal} {\bibinfo
  {journal} {Phys. Rev. Lett.}\ }\textbf {\bibinfo {volume} {112}},\ \bibinfo
  {pages} {050502} (\bibinfo {year} {2014})}\BibitemShut {NoStop}%
\bibitem [{\citenamefont {Shabani}\ \emph {et~al.}(2011)\citenamefont
  {Shabani}, \citenamefont {Kosut}, \citenamefont {Mohseni}, \citenamefont
  {Rabitz}, \citenamefont {Broome}, \citenamefont {Almeida}, \citenamefont
  {Fedrizzi},\ and\ \citenamefont {White}}]{simplifiedQPT1}%
  \BibitemOpen
  \bibfield  {author} {\bibinfo {author} {\bibfnamefont {A.}~\bibnamefont
  {Shabani}}, \bibinfo {author} {\bibfnamefont {R.~L.}\ \bibnamefont {Kosut}},
  \bibinfo {author} {\bibfnamefont {M.}~\bibnamefont {Mohseni}}, \bibinfo
  {author} {\bibfnamefont {H.}~\bibnamefont {Rabitz}}, \bibinfo {author}
  {\bibfnamefont {M.~A.}\ \bibnamefont {Broome}}, \bibinfo {author}
  {\bibfnamefont {M.~P.}\ \bibnamefont {Almeida}}, \bibinfo {author}
  {\bibfnamefont {A.}~\bibnamefont {Fedrizzi}}, \ and\ \bibinfo {author}
  {\bibfnamefont {A.~G.}\ \bibnamefont {White}},\ }\href {\doibase
  10.1103/PhysRevLett.106.100401} {\bibfield  {journal} {\bibinfo  {journal}
  {Phys. Rev. Lett.}\ }\textbf {\bibinfo {volume} {106}},\ \bibinfo {pages}
  {100401} (\bibinfo {year} {2011})}\BibitemShut {NoStop}%
\bibitem [{\citenamefont {Wu}\ \emph {et~al.}(2013)\citenamefont {Wu},
  \citenamefont {Li}, \citenamefont {Zheng}, \citenamefont {Peng},\ and\
  \citenamefont {Feng}}]{simplifiedQPT2}%
  \BibitemOpen
  \bibfield  {author} {\bibinfo {author} {\bibfnamefont {Z.}~\bibnamefont
  {Wu}}, \bibinfo {author} {\bibfnamefont {S.}~\bibnamefont {Li}}, \bibinfo
  {author} {\bibfnamefont {W.}~\bibnamefont {Zheng}}, \bibinfo {author}
  {\bibfnamefont {X.}~\bibnamefont {Peng}}, \ and\ \bibinfo {author}
  {\bibfnamefont {M.}~\bibnamefont {Feng}},\ }\href {\doibase
  http://dx.doi.org/10.1063/1.4774119} {\bibfield  {journal} {\bibinfo
  {journal} {The Journal of Chemical Physics}\ }\textbf {\bibinfo {volume}
  {138}},\ \bibinfo {pages} {024318} (\bibinfo {year} {2013})}\BibitemShut
  {NoStop}%
\bibitem [{\citenamefont {Mazzei}\ \emph {et~al.}(2003)\citenamefont {Mazzei},
  \citenamefont {Ricci}, \citenamefont {De~Martini},\ and\ \citenamefont
  {D'Ariano}}]{mazzei2003pauli}%
  \BibitemOpen
  \bibfield  {author} {\bibinfo {author} {\bibfnamefont {A.}~\bibnamefont
  {Mazzei}}, \bibinfo {author} {\bibfnamefont {M.}~\bibnamefont {Ricci}},
  \bibinfo {author} {\bibfnamefont {F.}~\bibnamefont {De~Martini}}, \ and\
  \bibinfo {author} {\bibfnamefont {G.}~\bibnamefont {D'Ariano}},\ }\href@noop
  {} {\bibfield  {journal} {\bibinfo  {journal} {Fortschritte der Physik}\
  }\textbf {\bibinfo {volume} {51}},\ \bibinfo {pages} {342} (\bibinfo {year}
  {2003})}\BibitemShut {NoStop}%
\bibitem [{\citenamefont {D'Ariano}\ and\ \citenamefont
  {Lo~Presti}(2003)}]{PRLAriano2003}%
  \BibitemOpen
  \bibfield  {author} {\bibinfo {author} {\bibfnamefont {G.~M.}\ \bibnamefont
  {D'Ariano}}\ and\ \bibinfo {author} {\bibfnamefont {P.}~\bibnamefont
  {Lo~Presti}},\ }\href {\doibase 10.1103/PhysRevLett.91.047902} {\bibfield
  {journal} {\bibinfo  {journal} {Phys. Rev. Lett.}\ }\textbf {\bibinfo
  {volume} {91}},\ \bibinfo {pages} {047902} (\bibinfo {year}
  {2003})}\BibitemShut {NoStop}%
\bibitem [{\citenamefont {Shukla}\ and\ \citenamefont
  {Mahesh}(2014)}]{shukla2014single}%
  \BibitemOpen
  \bibfield  {author} {\bibinfo {author} {\bibfnamefont {A.}~\bibnamefont
  {Shukla}}\ and\ \bibinfo {author} {\bibfnamefont {T.~S.}\ \bibnamefont
  {Mahesh}},\ }\href@noop {} {\bibfield  {journal} {\bibinfo  {journal}
  {Physical Review A}\ }\textbf {\bibinfo {volume} {90}},\ \bibinfo {pages}
  {052301} (\bibinfo {year} {2014})}\BibitemShut {NoStop}%
\bibitem [{\citenamefont {Meiboom}\ and\ \citenamefont {Gill}(1958)}]{cp}%
  \BibitemOpen
  \bibfield  {author} {\bibinfo {author} {\bibfnamefont {S.}~\bibnamefont
  {Meiboom}}\ and\ \bibinfo {author} {\bibfnamefont {D.}~\bibnamefont {Gill}},\
  }\href {\doibase http://dx.doi.org/10.1063/1.1716296} {\bibfield  {journal}
  {\bibinfo  {journal} {Rev. of Sci. Instrum.}\ }\textbf {\bibinfo {volume}
  {29}},\ \bibinfo {pages} {688} (\bibinfo {year} {1958})}\BibitemShut
  {NoStop}%
\bibitem [{\citenamefont {Viola}\ \emph {et~al.}(1999)\citenamefont {Viola},
  \citenamefont {Knill},\ and\ \citenamefont {Lloyd}}]{lloyd1}%
  \BibitemOpen
  \bibfield  {author} {\bibinfo {author} {\bibfnamefont {L.}~\bibnamefont
  {Viola}}, \bibinfo {author} {\bibfnamefont {E.}~\bibnamefont {Knill}}, \ and\
  \bibinfo {author} {\bibfnamefont {S.}~\bibnamefont {Lloyd}},\ }\href
  {\doibase 10.1103/PhysRevLett.82.2417} {\bibfield  {journal} {\bibinfo
  {journal} {Phys. Rev. Lett.}\ }\textbf {\bibinfo {volume} {82}},\ \bibinfo
  {pages} {2417} (\bibinfo {year} {1999})}\BibitemShut {NoStop}%
\bibitem [{\citenamefont {Viola}\ and\ \citenamefont {Lloyd}(1998)}]{lloyd2}%
  \BibitemOpen
  \bibfield  {author} {\bibinfo {author} {\bibfnamefont {L.}~\bibnamefont
  {Viola}}\ and\ \bibinfo {author} {\bibfnamefont {S.}~\bibnamefont {Lloyd}},\
  }\href {\doibase 10.1103/PhysRevA.58.2733} {\bibfield  {journal} {\bibinfo
  {journal} {Phys. Rev. A}\ }\textbf {\bibinfo {volume} {58}},\ \bibinfo
  {pages} {2733} (\bibinfo {year} {1998})}\BibitemShut {NoStop}%
\bibitem [{\citenamefont {Uhrig}(2007)}]{uhrig}%
  \BibitemOpen
  \bibfield  {author} {\bibinfo {author} {\bibfnamefont {G.~S.}\ \bibnamefont
  {Uhrig}},\ }\href {\doibase 10.1103/PhysRevLett.98.100504} {\bibfield
  {journal} {\bibinfo  {journal} {Phys. Rev. Lett.}\ }\textbf {\bibinfo
  {volume} {98}},\ \bibinfo {pages} {100504} (\bibinfo {year}
  {2007})}\BibitemShut {NoStop}%
\bibitem [{\citenamefont {Preskill}(1998)}]{preskill}%
  \BibitemOpen
  \bibfield  {author} {\bibinfo {author} {\bibfnamefont {J.}~\bibnamefont
  {Preskill}},\ }\href@noop {} {\bibfield  {journal} {\bibinfo  {journal}
  {Proc. R. Soc. Lond. A}\ }\textbf {\bibinfo {volume} {454}},\ \bibinfo
  {pages} {385} (\bibinfo {year} {1998})}\BibitemShut {NoStop}%
\bibitem [{\citenamefont {Farhi}\ \emph {et~al.}(2000)\citenamefont {Farhi},
  \citenamefont {Goldstone}, \citenamefont {Gutmann},\ and\ \citenamefont
  {Sipser}}]{farhi}%
  \BibitemOpen
  \bibfield  {author} {\bibinfo {author} {\bibfnamefont {E.}~\bibnamefont
  {Farhi}}, \bibinfo {author} {\bibfnamefont {J.}~\bibnamefont {Goldstone}},
  \bibinfo {author} {\bibfnamefont {S.}~\bibnamefont {Gutmann}}, \ and\
  \bibinfo {author} {\bibfnamefont {M.}~\bibnamefont {Sipser}},\ }\href@noop {}
  {\bibfield  {journal} {\bibinfo  {journal} {arXiv:quant-ph/0001106}\ }
  (\bibinfo {year} {2000})}\BibitemShut {NoStop}%
\bibitem [{\citenamefont {Lidar}\ and\ \citenamefont {Whaley}(2003)}]{DFS}%
  \BibitemOpen
  \bibfield  {author} {\bibinfo {author} {\bibfnamefont {D.~A.}\ \bibnamefont
  {Lidar}}\ and\ \bibinfo {author} {\bibfnamefont {K.~B.}\ \bibnamefont
  {Whaley}},\ }in\ \href@noop {} {\emph {\bibinfo {booktitle} {Irreversible
  Quantum Dynamics}}}\ (\bibinfo  {publisher} {Springer},\ \bibinfo {year}
  {2003})\ pp.\ \bibinfo {pages} {83--120}\BibitemShut {NoStop}%
\bibitem [{\citenamefont {Hegde}\ and\ \citenamefont
  {Mahesh}(2014)}]{hegde2014engineered}%
  \BibitemOpen
  \bibfield  {author} {\bibinfo {author} {\bibfnamefont {S.~S.}\ \bibnamefont
  {Hegde}}\ and\ \bibinfo {author} {\bibfnamefont {T.~S.}\ \bibnamefont
  {Mahesh}},\ }\href@noop {} {\bibfield  {journal} {\bibinfo  {journal}
  {Physical Review A}\ }\textbf {\bibinfo {volume} {89}},\ \bibinfo {pages}
  {062317} (\bibinfo {year} {2014})}\BibitemShut {NoStop}%
\bibitem [{\citenamefont {Cory}\ \emph {et~al.}(1998)\citenamefont {Cory},
  \citenamefont {Price},\ and\ \citenamefont {Havel}}]{cory}%
  \BibitemOpen
  \bibfield  {author} {\bibinfo {author} {\bibfnamefont {D.~G.}\ \bibnamefont
  {Cory}}, \bibinfo {author} {\bibfnamefont {M.~D.}\ \bibnamefont {Price}}, \
  and\ \bibinfo {author} {\bibfnamefont {T.~F.}\ \bibnamefont {Havel}},\ }\href
  {\doibase 10.1016/S0167-2789(98)00046-3} {\bibfield  {journal} {\bibinfo
  {journal} {Physica D}\ }\textbf {\bibinfo {volume} {120}},\ \bibinfo {pages}
  {82} (\bibinfo {year} {1998})}\BibitemShut {NoStop}%
\bibitem [{\citenamefont {Biercuk}\ \emph {et~al.}(2011)\citenamefont
  {Biercuk}, \citenamefont {Doherty},\ and\ \citenamefont {Uys}}]{biercuk}%
  \BibitemOpen
  \bibfield  {author} {\bibinfo {author} {\bibfnamefont {M.}~\bibnamefont
  {Biercuk}}, \bibinfo {author} {\bibfnamefont {A.}~\bibnamefont {Doherty}}, \
  and\ \bibinfo {author} {\bibfnamefont {H.}~\bibnamefont {Uys}},\ }\href@noop
  {} {\bibfield  {journal} {\bibinfo  {journal} {J. Phys. B}\ }\textbf
  {\bibinfo {volume} {44}},\ \bibinfo {pages} {154002} (\bibinfo {year}
  {2011})}\BibitemShut {NoStop}%
\bibitem [{\citenamefont {Biercuk}\ \emph {et~al.}(2009)\citenamefont
  {Biercuk}, \citenamefont {Uys}, \citenamefont {VanDevender}, \citenamefont
  {Shiga}, \citenamefont {Itano},\ and\ \citenamefont {Bollinger}}]{biercuk1}%
  \BibitemOpen
  \bibfield  {author} {\bibinfo {author} {\bibfnamefont {M.~J.}\ \bibnamefont
  {Biercuk}}, \bibinfo {author} {\bibfnamefont {H.}~\bibnamefont {Uys}},
  \bibinfo {author} {\bibfnamefont {A.~P.}\ \bibnamefont {VanDevender}},
  \bibinfo {author} {\bibfnamefont {N.}~\bibnamefont {Shiga}}, \bibinfo
  {author} {\bibfnamefont {W.~M.}\ \bibnamefont {Itano}}, \ and\ \bibinfo
  {author} {\bibfnamefont {J.~J.}\ \bibnamefont {Bollinger}},\ }\href@noop {}
  {\bibfield  {journal} {\bibinfo  {journal} {Nature}\ }\textbf {\bibinfo
  {volume} {458}},\ \bibinfo {pages} {996} (\bibinfo {year}
  {2009})}\BibitemShut {NoStop}%
\bibitem [{\citenamefont {Pan}\ \emph {et~al.}(2011)\citenamefont {Pan},
  \citenamefont {Xi},\ and\ \citenamefont {Gong}}]{pan}%
  \BibitemOpen
  \bibfield  {author} {\bibinfo {author} {\bibfnamefont {Y.}~\bibnamefont
  {Pan}}, \bibinfo {author} {\bibfnamefont {Z.-R.}\ \bibnamefont {Xi}}, \ and\
  \bibinfo {author} {\bibfnamefont {J.}~\bibnamefont {Gong}},\ }\href@noop {}
  {\bibfield  {journal} {\bibinfo  {journal} {J. Phys. B}\ }\textbf {\bibinfo
  {volume} {44}},\ \bibinfo {pages} {175501} (\bibinfo {year}
  {2011})}\BibitemShut {NoStop}%
\bibitem [{\citenamefont {Yuge}\ \emph {et~al.}(2011)\citenamefont {Yuge},
  \citenamefont {Sasaki},\ and\ \citenamefont {Hirayama}}]{yuge}%
  \BibitemOpen
  \bibfield  {author} {\bibinfo {author} {\bibfnamefont {T.}~\bibnamefont
  {Yuge}}, \bibinfo {author} {\bibfnamefont {S.}~\bibnamefont {Sasaki}}, \ and\
  \bibinfo {author} {\bibfnamefont {Y.}~\bibnamefont {Hirayama}},\ }\href@noop
  {} {\bibfield  {journal} {\bibinfo  {journal} {Phys. Rev. Lett.}\ }\textbf
  {\bibinfo {volume} {107}},\ \bibinfo {pages} {170504} (\bibinfo {year}
  {2011})}\BibitemShut {NoStop}%
\end{thebibliography}%
\end{document}